\documentclass[fleqn,twoside]{article}
\usepackage{espcrc2}

\usepackage{graphicx}
\usepackage{amssymb}
\usepackage[figuresright]{rotating}

\def\lsim{\hbox{ \raise.35ex\rlap{$<$}\lower.6ex\hbox{$\sim$}\ }}
\def\gsim{\hbox{ \raise.35ex\rlap{$>$}\lower.6ex\hbox{$\sim$}\ }}

\def\xrightarrow#1#2#3#4{\,\lower#1pt\hbox{$\stackrel{\stackrel{\displaystyle #2}%
{\hbox to #3cm{\rightarrowfill}}}{#4}$}\,}

\newcommand{\AmS}{{\protect\the\textfont2
  A\kern-.1667em\lower.5ex\hbox{M}\kern-.125emS}}

\title{Cosmic Strings and Cosmic Superstrings}

\author{ Mairi Sakellariadou \address{Department of Physics, King's
    College, University of London, Strand, London WC2R 2LS, U.K.} }
\begin{document}

\begin{abstract}
In these lectures, I review the current status of cosmic strings and
cosmic superstrings. I first discuss topological defects in the
context of Grand Unified Theories, focusing in particular in cosmic
strings arising as gauge theory solitons. I discuss the reconciliation
between cosmic strings and cosmological inflation, I review cosmic
string dynamics, cosmic string thermodynamics and cosmic string
gravity, which leads to a number of interesting observational
signatures. I then proceed with the notion of cosmic superstrings
arising at the end of brane inflation, within the context of
brane-world cosmological models inspired from string theory. I discuss
the differences between cosmic superstrings and their solitonic analogues,
I review our current understanding about the evolution of cosmic
superstring networks, and I then briefly describe the variety of
observational consequences, which may help us to get an insight into
the stringy description of our Universe.

\vspace{1pc}
\end{abstract}

\maketitle

\section{Introduction}
Provided our understanding about unification of forces and big bang
cosmology are correct, it is natural to expect that topological
defects, appearing as solutions to many particle physics models of
matter, could have formed naturally during phase transitions followed
by spontaneously broken symmetries, in the early stages of the
evolution of the Universe. Certain types of topological defects (local
monopoles and local domain walls) may lead to disastrous consequences
for cosmology, hence being undesired, while others (cosmic strings)
may play a useful r\^ole.

Cosmic strings~\cite{vs} are linear topological defects, analogous to
flux tubes in type-II superconductors, or to vortex filaments in
superfluid helium. These objects gained a lot of interest in the
1980's and early 1990's, since they offered a potential alternative to
the cosmological inflation for the origin of initial density
fluctuations leading to the Cosmic Microwave Background (CMB)
temperature anisotropies and the observed structure in the
Universe. They however lost their appeal, when it was found that they
lead to inconsistencies in the power spectrum of the CMB.  It was
later shown~\cite{Jeannerot:2003qv} that cosmic strings are
generically formed at the end of an inflationary era, within the
framework of Supersymmetric Grand Unified Theories (SUSY GUTs). Hence
cosmic strings have to be included as a sub-dominant partner of
inflation.  This theoretical support gave a new boost to the field of
cosmic strings, a boost which has been more recently enhanced when it
was shown that cosmic superstrings~\cite{Polchinski:2004ia}
(fundamental or one-dimensional Dirichlet branes) can play the r\^ole
of cosmic strings, in the framework of brane-world cosmologies.

A realistic cosmological scenario necessitates the input of high
energy physics; any models describing the early stages of the
evolution of the Universe have their foundations in general relativity
and high energy physics. Comparing the theoretical predictions of such
models against current astrophysical and cosmological data, results to
either their acceptance or their rejection, while in the first case it
also fixes the free parameters of the models ({\sl see}
e.g., Ref.~\cite{Rocher:2004et,Rocher:2004my}).  In particular, by
studying the properties of cosmic superstring networks and comparing
their phenomenological consequences against observational data, we
expect to pin down the successful and natural inflationary model and
get some insight into the stringy description of the Universe. Cosmic
strings/superstrings represent a beautiful example of the strong and
fruitful link between cosmology and high energy physics.

In what follows, I will summarise the material I had presented in my
lectures during the summer school at Carg\`ese (June 2008)
\footnote{http://www.lpthe.jussieu.fr/cargese/}. I will highlight only
certain aspects of the subject, which I consider either more important
due to their observational consequences, or more recently obtained
results.

\section{Topological Defects in GUTs}
In the framework of the hot big bang cosmological model, the Universe
was originally at a very high temperature, hence the initial
equilibrium value of the Higgs field $\phi$, which plays the r\^ole of
the order parameter, was at $\phi=0$.  Since the Planck time, the
Universe has, through its expansion, steadily cooled down and a series
of phase transitions followed by Spontaneously Symmetry
Breaking \footnote{The concept of spontaneous symmetry breaking has
  its origin in condensed matter physics.} (SSBs) took place in the
framework of GUTs. Such SSBs may have left behind topological defects
as false vacuum remnants, via the Kibble mechanism~\cite{kibble}.  

The formation or not of topological defects and the determination of
their type, depend on the topology of the vacuum manifold ${\cal
  M}_n$. The properties of ${\cal M}_n$ are described by the $k^{\rm
  th}$ homotopy group $\pi_k({\cal M}_n)$, which classifies distinct
mappings from the $k$-dimensional sphere $S^k$ into the manifold
${\cal M}_n$.  Consider the symmetry breaking of a group G down to a
subgroup H of G. If ${\cal M}_n = {\rm G}/{\rm H}$ has disconnected
components --- equivalently, if the order $k$ of the non-trivial
homotopy group is $k = 0$ --- two-dimensional defects, called domain
walls, form. The space-time dimension $d$ of the defects is given in
terms of the order of the non-trivial homotopy group by $d = 4-1-k$. If
${\cal M}_n$ is not simply connected --- equivalently, if ${\cal M}_n$
contains loops which cannot be continuously shrunk into a point ---
cosmic strings form. A necessary, but not sufficient, condition for
the existence of stable strings is that the fundamental group
$\pi_1$ of ${\cal M}_n$, is non-trivial, or multiply
connected. Cosmic strings are linear-like defects, $d = 2$. If ${\cal
  M}_n$ contains unshrinkable surfaces, then monopoles form; $k = 1, d
= 1$. If ${\cal M}_n$ contains non-contractible three-spheres, then
event-like defects, textures, form; $k = 3, d = 0$.  

Depending on whether the symmetry is local (gauged) or global (rigid),
topological defects are respectively, local or global. The energy of
local defects is strongly confined, while the gradient energy of
global defects is spread out over the causal horizon at defect
formation. Global defects having long range density fields and forces,
can decay through long-range interactions, hence they do not
contradict observations, while local defects may be undesirable for
cosmology.  In what follows, I will discuss local defects, since we
are interested in gauge theories, being the more physical
ones \footnote{Note that when we say cosmic strings we refer to local
  one-dimensional topological defects.}.  Patterns of symmetry
breaking which lead to the formation of local monopoles or local
domain walls are ruled out, since they should soon dominate the energy
density of the Universe and close it, unless an inflationary era took
place after their formation. This is one of the reasons for which
cosmological inflation --- a period in the earliest stages of the
evolution of the Universe, during which the Universe could be in an
unstable vacuum-like state having high energy density, which remained
almost constant --- was proposed. Local textures are insignificant in
cosmology since their relative contribution to the energy density of
the Universe decreases rapidly with time.

Even in the absence of a non-trivial topology in a field theory, it may
still be possible to have defect-like solutions, since defects may be
embedded in such topologically trivial field theories.  However, while
stability of topological defects is guaranteed by topology, embedded
defects are in general unstable under small perturbations.

Let me discuss the genericity of cosmic string formation in the
context of SUSY GUTs, which contain a large number of SSB patterns
leading from a large gauge group G$_{\rm GUT}$ to the Standard Model
(SM) gauge group G$_{\rm SM}\equiv$ SU(3)$_{\rm C}\times$ SU(2)$_{\rm
L}\times$ U(1)$_{\rm Y}$. The minimum rank of G$_{\rm GUT}$ has to be
at least equal to 4, to contain the G$_{\rm SM}$ as a subgroup; we set
the upper bound on the rank $r$ of the group to be $r\leq 8$.  The
embeddings of G$_{\rm SM}$ in G$_{\rm GUT}$ must be such that there is
an agreement with the SM phenomenology and especially with the
hypercharges of the known particles. The large gauge group G$_{\rm
GUT}$ must include a complex representation, needed to describe the SM
fermions, and it must be anomaly free.  A detailed
investigation~\cite{Jeannerot:2003qv} has concluded that G$_{\rm GUT}$
could be either one of SO(10), E$_6$, SO(14), SU(8), SU(9); flipped
SU(5) and [SU(3)]$^3$ are included within this list as subgroups of
SO(10) and E$_6$, respectively.  The formation of domain walls or
monopoles, necessitates an era of supersymmetric hybrid inflation to
dilute them. Considering GUTs based on simple gauge groups, the type
of supersymmetric hybrid inflation will be of the F-type.  The
baryogenesis mechanism will be obtained via leptogenesis, either
thermal or non-thermal leptogenesis. Finally, to ensure the stability
of proton, the discrete symmetry Z$_2$, which is contained in
U(1)$_{\rm B-L}$, must be kept unbroken down to low energies; the
successful SSB schemes should end at G$_{\rm SM}\times$ Z$_2$. Taking
all these considerations into account, a detailed study of all SSB
schemes leading from a G$_{\rm GUT}$ down to the G$_{\rm SM}$, by one
or more intermediate steps, shows that cosmic strings are generically
formed at the end of hybrid inflation.

The results~\cite{Jeannerot:2003qv} can be summarised as follows: If
the large gauge group G$_{\rm GUT}$ is the SO(10), then cosmic strings
formation is unavoidable.  The genericity of string formation in the
case that the large gauge group is the E$_6$, depends upon whether one
considers thermal or non-thermal leptogenesis. More precisely, for
non-thermal leptogenesis, cosmic string formation is unavoidable,
while for thermal leptogenesis, cosmic string formation arises in
98$\%$ of the acceptable SSB schemes. If the requirement of having
Z$_2$ unbroken down to low energies is relaxed and thermal
leptogenesis is considered as being the mechanism for baryogenesis,
then cosmic string formation accompanies hybrid inflation in 80$\%$
of the SSB schemes. The SSB schemes of either SU(6) or SU(7), as the
large gauge group, down to the G$_{\rm SM}$, which could accommodate
an inflationary era with no defect (of any kind) at later times are
inconsistent with proton lifetime measurements, while minimal SU(6)
and SU(7) do not predict neutrino masses, implying that these models
are incompatible with high energy physics phenomenology. Higher rank
groups, namely SO(14), SU(8) and SU(9), should in general lead to
cosmic string formation at the end of hybrid inflation. In all these
schemes, cosmic string formation is sometimes accompanied by the
formation of embedded strings.  The strings which form at the end of
hybrid inflation have a mass which is proportional to the inflationary
scale.

\section{Cosmic Strings and Inflation}
An appealing solution to the drawbacks of the standard hot big bang
model is to introduce, during the very early stages of the evolution
of the Universe, a period of accelerated expansion, known as
cosmological inflation~\cite{inflation}.  The inflationary era took
place when the Universe was in an unstable vacuum-like state at a high
energy density, leading to a quasi-exponential expansion. The
combination of the hot big bang model and the inflationary scenario
provides at present the most comprehensive picture of the Universe at
our disposal. Inflation ends when the Hubble parameter $H =\sqrt{
  8\pi\rho/(3M_{\rm Pl}^2)}$ (where $\rho$ denotes the energy density
and $M_{\rm Pl}$ stands for the Planck mass) starts decreasing rapidly. The
energy stored in the vacuum-like state gets transformed into thermal
energy, heating up the Universe and leading to the beginning of the
standard hot big bang radiation-dominated era.

Inflation is based on the basic principles of general relativity and
field theory, while when the principles of quantum mechanics are also
considered, it provides a successful explanation for the origin of the
large scale structure, associated with the measured temperature
anisotropies in the CMB spectrum. Despite its remarkable success,
inflation still remains a paradigm in search of model.  An
inflationary model should be inspired from a fundamental theory, while
its predictions should be tested against current data. In addition,
releasing the present Universe form its acute dependence on the
initial data, inflation is faced with the challenging task of proving
itself generic~\cite{Calzetta:1992gv}, in the sense that inflation
would take place without fine-tuning of the initial conditions.

Theoretically motivated inflationary models can be built in the
context of supersymmetry or Supergravity (SUGRA). N=1
supersymmetry models contain complex scalar fields which often have
flat directions in their potential, thus offering natural candidates
for inflationary models. In this framework, hybrid inflation driven by
F-terms or D-terms is the standard inflationary model, leading
generically to cosmic string formation at the end of inflation.
Hybrid inflation is based on Einstein's gravity but is driven by the
false vacuum. The inflaton field rolls down its potential while
another scalar field is trapped in an unstable false vacuum. Once the
inflaton field becomes much smaller than some critical value, a phase
transition to the true vacuum takes place and inflation ends.  F-term
inflation is potentially plagued with the {\sl Hubble-induced mass}
problem \footnote{In supergravity theories, the supersymmetry
breaking is transmitted to all fields by gravity, and thus any scalar
field, including the inflaton, gets an effective mass of the order of
the expansion rate $H$ during inflation.} ($\eta$-problem), while
D-term inflation avoids it.

F-term inflation can be naturally accommodated in the framework of
GUTs, when a G$_{\rm GUT}$ is broken down to the G$_{\rm SM}$, at an
energy scale $M_{\rm GUT}$ according to the scheme
$${\rm G}_{\rm GUT} \stackrel{M_{\rm GUT}}{\hbox to 0.8cm
{\rightarrowfill}} {\rm H}_1 \xrightarrow{9}{M_{\rm
infl}}{1}{\Phi_+\Phi_-} {\rm H}_2 {\longrightarrow} {\rm G}_{\rm SM}~,$$
where $\Phi_+, \Phi_-$ is a pair of GUT Higgs superfields in
non-trivial complex conjugate representations, which lower the rank of
the group by one unit when acquiring non-zero vacuum expectation
value. The inflationary phase takes place at the beginning of the
symmetry breaking ${\rm H}_1\stackrel{M_{\rm infl}}{\longrightarrow}
{\rm H}_2$.  The gauge symmetry is spontaneously broken by adding
F-terms to the superpotential. The Higgs mechanism leads
generically~\cite{Jeannerot:2003qv} to Abrikosov-Nielsen-Olesen
strings, called F-term strings.

F-term inflation is based on the globally supersymmetric
renormalisable superpotential
\begin{equation}\label{superpot}
W_{\rm infl}^{\rm F}=\kappa  S(\Phi_+\Phi_- - M^2)~,
\end{equation}
where $S$ is a GUT gauge singlet left handed superfield and $\kappa$, $M$
 are two constants ($M$ has dimensions of mass) which can be taken
 positive with field redefinition.

The scalar potential, as a function of the scalar complex component
 of the respective chiral superfields $\Phi_\pm, S$, is
\begin{eqnarray}
\label{scalpot1}
V(\phi_+,\phi_-, S)= |F_{\Phi_+}|^2+|F_{\Phi_-}|^2+|F_ S|^2\nonumber\\
\ \ \ \ \   \ \ \ \ \ \ \ \ \ \ \ \ \ +\frac{1}{2}\sum_a g_a^2 D_a^2~.
\end{eqnarray}
The F-term is such that $F_{\Phi_i} \equiv |\partial W/\partial
\Phi_i|_{\theta=0}$, where we take the scalar component of the
superfields once we differentiate with respect to $\Phi_i=\Phi_\pm,
S$. The D-terms are $D_a=\bar{\phi}_i\,{(T_a)^i}_j\,\phi^j +\xi_a$,
with $a$ the label of the gauge group generators $T_a$, $g_a$ the
gauge coupling, and $\xi_a$ the Fayet-Iliopoulos term. By definition,
in the F-term inflation the real constant $\xi_a$ is zero; it can only
be nonzero if $T_a$ generates an extra U(1) group.  In the context of
F-term hybrid inflation the F-terms give rise to the inflationary
potential energy density while the D-terms are flat along the
inflationary trajectory, thus one may neglect them during inflation.

The potential, has one valley of local minima, $V=\kappa^2 M^4$, for
$S> M $ with $\phi_+ = \phi_-=0$, and one global supersymmetric
minimum, $V=0$, at $S=0$ and $\phi_+ = \phi_- = M$. Imposing initially
$ S \gg M$, the fields quickly settle down the valley of local minima.
Since in the slow-roll inflationary valley the ground state of the
scalar potential is non-zero, supersymmetry is broken.  In the tree
level, along the inflationary valley the potential is constant,
therefore perfectly flat. A slope along the potential can be generated
by including one-loop radiative corrections.  Hence, the scalar
potential gets a little tilt which helps the inflaton field $S$ to
slowly roll down the valley of minima. The one-loop radiative
corrections to the scalar potential along the inflationary valley lead
to the effective potential~\cite{Rocher:2004et}
\begin{eqnarray}
\label{VexactF}
V_{\rm eff}^{\rm F}(|S|)&=&\kappa^2M^4\biggl\{1+\frac{\kappa^2
\cal{N}}{32\pi^2}\biggl[2\ln\frac{|S|^2\kappa^2}{\Lambda^2}\nonumber\\
&&~~~~~~~+(z+1)^2
\ln(1+z^{-1})\nonumber\\
&&~~~~~~~
+(z-1)^2\ln(1-z^{-1})
\biggr]\biggr\},
\end{eqnarray}
with $z=|S|^2/M^2$, and $\cal{N}$ stands for the dimensionality of the
representation to which the complex scalar components $\phi_+, \phi_-$
of the chiral superfields $\Phi_+, \Phi_-$ belong. This implies that
the effective potential, Eq.~(\ref{VexactF}), depends on the
particular symmetry breaking scheme considered.

D-term inflation can be easily implemented within high energy physics
(e.g., SUSY GUTs, SUGRA, or string theories) and it avoids the
$\eta$-problem. Within D-term inflation, the gauge symmetry is
spontaneously broken by introducing Fayet-Iliopoulos (FI) D-terms. In
standard D-term inflation, the constant FI term gets compensated by a
single complex scalar field at the end of the inflationary era, which
implies that standard D-term inflation ends always with the formation
of cosmic strings, called D-term strings.  A supersymmetric description of
the standard D-term inflation is insufficient, since the inflaton
field reaches values of the order of the Planck mass, or above it,
even if one concentrates only around the last 60 e-folds of
inflation. Thus, D-term inflation has to be studied in the context of
supergravity~\cite{Rocher:2004et,Rocher:2004my}).

Standard D-term inflation requires a scheme
$${\rm G}_{\rm GUT}\times {\rm U}(1) \stackrel{M_{\rm GUT}}{\hbox to
  0.8cm{\rightarrowfill}} {\rm H} \times {\rm U}(1)
\xrightarrow{9}{M_{\rm infl}}{1}{\Phi_+\Phi_-} {\rm H} \rightarrow {\rm
  G}_{\rm SM}~.
$$
It is based on the superpotential
\begin{equation}\label{superpoteninflaD}
W=\lambda S\Phi_+\Phi_-~,
\end{equation}
where $S, \Phi_+, \Phi_-$ are three chiral superfields and $\lambda$
is the superpotential coupling. It assumes an invariance under an
Abelian gauge group $U(1)_\xi$, under which the superfields $S,
\Phi_+, \Phi_-$ have charges $0$, $+1$ and $-1$, respectively. It also
assumes the existence of a constant FI term $\xi$.  In D-term
inflation the superpotential vanishes at the unstable de Sitter vacuum
(anywhere else the superpotential is non-zero), implying that when the
superpotential vanishes, D-term inflation must be studied within a
non-singular formulation of supergravity.  Various formulations of
effective supergravity can be constructed from the superconformal
field theory.  To construct a formulation of supergravity with
constant FI terms from superconformal theory, one
finds~\cite{Binetruy:2004hh} that under U(1) gauge transformations in
the directions in which there are constant FI terms $\xi_\alpha$, the
superpotential $W$ must transform as $\delta_\alpha W=\eta_{\alpha
i}\partial^i W = -i (g\xi_\alpha/M_{\rm Pl}^2)W$; one cannot keep any
longer the same charge assignments as in {\sl standard}
supergravity. 

D-term inflationary models can be built with
different choices of the K\"ahler geometry. Various cases have been
explored in the literature. The simplest case is that of D-term
inflation within minimal supergravity~\cite{Rocher:2004et}. It is
based on
\begin{equation}\label{Kmin}
K_{\rm min}=\sum_i |\Phi_i|^2=|\Phi_-|^2+|\Phi_+|^2+|S|^2~,
\end{equation}
with $f_{ab}(\Phi_i)=\delta_{ab}$.  

Another example is that of D-term inflation based on K\"ahler geometry
with shift symmetry,
\begin{equation}\label{K3}
K_{\rm shift}=\frac{1}{2} (S+\bar{S})^2+|\phi_+|^2+|\phi_-|^2~,
\end{equation}
and minimal structure for the kinetic function~\cite{Rocher:2004my}.

One can also consider consider~\cite{Rocher:2004my} a K\"ahler
potential with non-renormalisable terms:
\begin{eqnarray}
K_{\rm
non-renorm}&=&|S|^2+|\Phi_+|^2+|\Phi_-|^2\nonumber\\
&& +f_+\bigg(\frac{|S|^2}{M_{\rm
Pl}^2}\bigg)|\Phi_+|^2\nonumber\\
&&+f_-\bigg(\frac{|S|^2}{M_{\rm Pl}^2}\bigg)
|\Phi_-|^2+b\frac{|S|^4}{M_{\rm Pl}^2},
\label{gen}
\end{eqnarray}
where $f_\pm$ are arbitrary functions of $(|S|^2/M_{\rm Pl}^2)$ and
the superpotential is given in Eq.~(\ref{superpoteninflaD}). 

Having the superpotential, one must proceed in the same way as in
F-term inflation and write down the three level scalar potential and
then include the one-loop radiate corrections.

Let me finally note that different approaches~\cite{Lazarides:1995vr}
have been proposed in order to avoid cosmic string formation in the
context of D-term inflation. For example, one can add a
non-renormalisable term in the potential, or add an additional discrete
symmetry, or consider GUT models based on non-simple groups, or finally
introduce a new pair of charged superfields so that cosmic string
formation is avoided within D-term inflation.

\section{Cosmic String Dynamics}

The world history of a cosmic string can be expressed by a
two-dimensional surface in the four-dimensional string world-sheet:
\begin{equation}
x^\mu=x^\mu(\zeta^a)\;,~~a=0,1~;
\end{equation}
the world-sheet coordinates $\zeta^0, \zeta^1$ are arbitrary
parameters, $\zeta^0$ is time-like and $\zeta^1$ ($\equiv
\sigma$) is space-like.

Over distances that are large compared to the width of the string, but
small compared to the horizon size, solitonic cosmic strings can be
considered as one-dimensional objects and their motion can be
well-described by the Nambu-Goto action.  Thus, the string equation
of motion, in the limit of a zero thickness string, is derived from
the Goto-Nambu effective action
\begin{equation}
S_0[x^\mu]=-\mu\int\sqrt{-\gamma}d^2\zeta\;,
\label{nga}
\end{equation}
where $\gamma={\rm det}(\gamma_{ab})$ with
$\gamma_{ab}=g_{\mu\nu}x^\mu_{,a}x^\nu_{,b}$ and $\mu$ stands for the
linear mass density, with $\mu\sim T_{\rm c}^2$, where $T_{\rm c}$ is
the critical temperature of the phase transition followed by SSB
leading to cosmic string formation.
By varying the action, Eq.~(\ref{nga}), with respect to
$x^\mu(\zeta^a)$, and using $d\gamma=\gamma \gamma^{ab}d\gamma_{ab}$,
we get the string equation of motion:
\begin{equation}
x^{\mu\
;a}_{,a}+\Gamma^\mu_{\nu\sigma}\gamma^{ab}x^\nu_{,a}x^\sigma_{,b}=0\;~;
\label{semnb}
\end{equation}
$\Gamma^\mu_{\nu\sigma}$ is the four-dimensional Christoffel
symbol.

We have neglected the friction~\cite{friction}, due to the scattering
of thermal particles off the string.  For strings formed at the grand
unification scale, friction is important only for a very short period
of time. For strings formed at a later phase transition (e.g., closer
to the electroweak scale), friction would dominate their dynamics
through most of the thermal history of the Universe.

By varying the action with respect to the metric, the string
energy-momentum tensor reads
$$T^{\mu\nu}\sqrt{-g}
=\mu\int
d^2\zeta\sqrt{-\gamma}\gamma^{ab}x^\mu_{,a}x^\nu_{,b}
\delta^{(4)}(x^\sigma-x^\sigma(\zeta^a))\;.
$$

In an expanding Universe, the cosmic string equation of motion is
most conveniently written in comoving coordinates, where the
Friedmann-Lema\^{i}tre-Robertson-Walker (FLRW) metric takes the form
${\rm d}s^2=a^2(\tau)[{\rm d}\tau^2-{\rm d}{\bf r}^2]\;;$
$a(\tau)$ is the cosmic scale factor in terms of conformal time $\tau$
(related to cosmological time $t$, by $dt=ad\tau$).  Under the gauge
condition $\zeta^0=\tau$, the comoving spatial string coordinates,
${\bf x}(\tau,\sigma)$, are written as a function of $\tau$, and the
length parameter $\sigma$. For a string moving in a FLRW Universe, the
equation of motion, Eq.~(\ref{semnb}), can be simplified in the gauge
for which the unphysical parallel components of the velocity vanish,
\begin{equation}
\dot {\bf x}\cdot {\bf x}'=0\;;
\label{gce}
\end{equation}
overdots and primes denote derivatives with respect to $\tau$ and
$\sigma$, respectively.  In these coordinates, the Goto-Nambu action
yields the following string equation of motion in a FLRW metric:
\begin{equation}
\ddot {\bf x}+2\biggl(\frac{\dot a}{a}\biggr)\dot {\bf x} (1-\dot {\bf
x}^2)=\biggl(\frac{1}{\epsilon}\biggr)\biggl(\frac{{\bf
x}'}{\epsilon}\biggr)'\;.
\label{eomflrw}
\end{equation}
The string energy per unit $\sigma$, in comoving units, is
$\epsilon\equiv\sqrt{{\bf x}^{'2}/(1-\dot{\bf x}^2)}$, implying that
the string energy is $\mu a \int \epsilon d\sigma$.  
One usually fixes entirely the gauge by choosing $\sigma$ so that
$\epsilon=1$ initially.\\

The string equation of motion is much simpler in Minkowski
space-time.  Equation (\ref{semnb}) for flat space-time simplifies to
\begin{equation}
\partial_a(\sqrt{-\gamma}\gamma^{ab}x^\mu_{,b})=0\;.
\end{equation}
We impose the conformal gauge
\begin{equation}
\dot x\cdot x'=0~ ~ ,~ ~ \dot x^2+x'^2=0\;;
\label{cem}
\end{equation}
overdots and primes denote derivatives with respect to $\zeta^0$ and
$\zeta^1$, respectively.  In this gauge the string equation of motion
is just a two-dimensional wave equation:
\begin{equation}
\ddot{\bf x}-{\bf x}''=0\;.
\label{eomm}
\end{equation}
To fix entirely the gauge, we set $t\equiv x^0=\zeta^0\;,$ which
allows us to write the string trajectory as the three-dimensional
vector ${\bf x}(\sigma,t)$, where $\zeta^1\equiv\sigma$, the space-like
parameter along the string. Hence, the constraint equations,
Eq.~(\ref{cem}), and the string equation of motion, Eq.~(\ref{eomm}),
become
\begin{equation}
\dot{\bf x}\cdot{\bf x}'=0\ \ ,\ \ \dot{{\bf x}}^2+{\bf x}'^2 =1\ \ ,
\ \ \ddot{\bf x}-{\bf x}'' =0\;.
\label{semf}
\end{equation}
The above equations imply that the string moves perpendicularly to
itself with velocity $\dot{\bf x}$, that $\sigma$ is proportional to
the string energy, and that the string acceleration in the string rest
frame is inversely proportional to the local string curvature radius.
A curved string segment tends to straighten itself, resulting to
string oscillations.

The general solution to the string equation of motion in flat
space-time, Eq.~(\ref{semf}c), is
\begin{equation}
{\bf x}=\frac{1}{2} [{\bf a}(\sigma-t)+{\bf
b}(\sigma+t) ]\;,
\end{equation}
where ${\bf a}(\sigma-t)$ and ${\bf b}(\sigma+t)$ are two
continuous arbitrary functions which satisfy
\begin{equation}
{\bf a}'^2={\bf b}'^2=1\;.
\end{equation}
Hence, $\sigma$ is the length parameter along the three-dimensional
curves ${\bf a}(\sigma), {\bf b}(\sigma)$.

The Goto-Nambu action describes to a good approximation cosmic string
segments which are separated. However, it leaves unanswered the issue
of what happens when strings cross; a study which necessitates full
field theory. When two strings of the same type collide, they may
either pass simply through one another, or they may reconnect
(intercommute). A necessary, but not sufficient, condition for string
reconnection is that the initial and final configurations be
kinematically allowed in the infinitely thing string approximation.
Numerical simulations (and analytical estimates) of type-II (and
weakly type-I) strings in the Abelian Higgs model suggest that the
probability that a pair of strings will reconnect, after they
intersect, is close to unity. The results are based on lattice
simulations of the corresponding classical field configurations in the
Abelian Higgs model; the internal structure of strings is highly
non-linear, and thus difficult to treat via analytical means.
String-string and self-string intersections lead to the formation of
new long strings and loops.  String intercommutations produce
discontinuities, {\sl kinks}, in $\dot{\bf x}$ and ${\bf x}'$ on the
new string segments at the intersection point, composed of right- and
left-moving pieces travelling along the string at the speed of light.

The first analytical studies of the evolution of a cosmic string
network have shown~\cite{one-scale} the existence of {\sl scaling}, in
the sense that the string network can be characterised by a single
length scale, roughly the persistence length or the inter-string
distance $\xi$ which grows with the horizon.  This important property
of cosmic strings renders them cosmologically acceptable, in contrast
to local monopoles or domain walls.  Early numerical simulations
have shown~\cite{numcs1} that the typical curvature radius of long
strings and the characteristic distance between the strings are both
comparable to the evolution time $t$.  The energy density of
super-horizon \footnote{Often in the literature, strings are divided
into two classes: string loops and {\sl infinite}, or long,
strings. However, in numerical simulations strings are always loops,
in the sense that they do not have open ends. Hence, the term {\sl
loops} corresponds to sub-horizon string loops, while the term {\sl
infinite strings} corresponds to super-horizon string loops.} strings
in the scaling regime is given (in the radiation-dominated era) by
$\rho_{\rm long} = \kappa\mu t^{-2} ~,$ where $\kappa$ is a numerical
coefficient $(\kappa=20\pm 10)$.  Assuming that the super-horizon
strings are characterised by a single length scale $\xi(t)$, implies
\begin{equation}
\xi(t)
=\kappa^{-1/2} t~.
\end{equation}
The typical distance between the nearest string segments and the
typical curvature radius of the strings are both of the order of
$\xi$.  These results agree with the picture of the scale-invariant
evolution of the string network and with the one-scale hypothesis.
Further numerical investigations however revealed dynamical processes,
such as the production of small sub-horizon loops, at scales much
smaller than $\xi$~\cite{numcs2}.  In response to these findings, a
three-scale model was developed~\cite{Austin:1993rg} which describes
the network in therms of three scales: the energy density scale $\xi$,
a correlation length $\bar\xi$ along the string, and a scale $\zeta$
related to local structure on the string. The small-scale structure
(wiggliness), which offers an explanation for the formation of the
small sub-horizon sized loops, is basically developed through
intersections of long string segments.  Aspects of the three-scale
model have been checked~\cite{Vincent:1996rb} evolving a cosmic string
network in Minkowski space-time.

\begin{figure}[t]
\centerline{\includegraphics[width=2.4in]
 {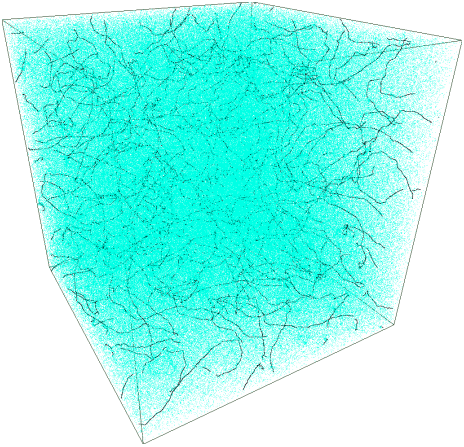}}
\caption{Snapshot of a network of long strings and closed loops in the
matter-dominated era. Figure taken from Ref.~\cite{Ringeval:2005kr}.}
\label{simulFig}
\end{figure}

The sub-horizon strings (loops), their size distribution, and the
mechanism of their formation remained for years the least understood
parts of the string evolution. Recently, numerical simulations of
cosmic string evolution in a FLRW Universe ({\sl see},
Fig.~\ref{simulFig}), found~\cite{Ringeval:2005kr} evidence of a
scaling regime for the cosmic string loops in the radiation- and
matter-dominated eras down to the hundredth of the horizon time. The
scaling was found without considering any gravitational back reaction
effect; it was just the result of string intercommutations. The scaling
regime of string loops appears after a transient relaxation era,
driven by a transient overproduction of string loops with length close
to the initial correlation length of the string network. Subsequently,
numerical~\cite{Vanchurin:2005pa} and
analytical~\cite{Polchinski:2006ee} studies supported the results of
Ref.~\cite{Ringeval:2005kr}.

Let me note that there are two approaches of developing numerical
simulations of cosmic string evolution. Either cosmic strings are
modelled as idealised one-dimensional objects, or field theoretic
calculations have been considered. In particular, for the field
theoretic approach, the simplest example of an underlying field theory
containing local U(1) strings, namely the Abelian Higgs model, has
been recently employed~\cite{hsb}.

\section{String Thermodynamics}
It is well-known in string theory, that the degeneracy of string states
increases exponentially with energy, namely
\begin{equation}
{\rm d}(E)\sim e^{\beta_{\rm H}E}~.
\end{equation}
Hence, there is a maximum temperature $T_{\rm max}=1/\beta_{\rm H}$,
the Hagedorn temperature~\cite{hagedorn}. Let us consider, in the
microcanonical ensemble, a system of closed string loops in a
three-dimensional box.  Intersecting strings intercommute, but
otherwise they do not interact and are described by the Goto-Nambu
equation of motion. The statistical properties of a system of strings
in equilibrium are characterised by only one parameter, the energy
density of strings, $\rho$, defined as $\rho=E/L^3$, with $L$ the size
of the cubical box.  The behaviour of the system depends on whether it
is at low or high energy densities, and it undergoes a phase
transition at a critical energy density, the Hagedorn energy density
$\rho_{\rm H}$.  Quantisation implies a lower cutoff for the size of
the string loops, determined by the string tension $\mu$. The lower
cutoff on the loop size is roughly $\mu^{-1/2}$, implying that the
mass of the smallest string loops is $m_0\sim \mu^{1/2}$.

For a system of strings at the low energy density regime
($\rho\ll\rho_{\rm H}$), all strings are chopped down to the loops of
the smallest size, while larger loops are exponentially suppressed.
Thus, for small enough energy densities, the string equilibrium
configuration is dominated by the massless modes in the quantum
description.  The energy distribution of loops, given by the number
${\rm d}n$ of loops with energies between $E$ and $E+{\rm d}E$ per
unit volume, is~\cite{hagedorn,ed3}
\begin{equation}
{\rm d}n\propto e^{-\alpha E} E^{-5/2} {\rm d}E ~~(\rho\ll\rho_{\rm H})~,
\end{equation}
where $\alpha=(5/2m_0)\ln(\rho_{\rm H}/\rho)$. 

However, as the energy density increases, more and more oscillatory
modes of strings get excited. In particular, once a critical energy
density, $\rho_{\rm H}$, is reached, long oscillatory string states
begin to appear in the equilibrium state. The density at which this
happens corresponds to the Hagedorn temperature. The Hagedorn energy
density --- achieved when the separation between the smallest string
loops is of the order of their sizes --- is approximately $\mu^2$, and
then the system undergoes a phase transition characterised by the
appearance of super-horizon (infintely long) strings.

At the high energy density regime, the energy
distribution of string loops is~\cite{hagedorn,ed3}
\begin{equation}
{\rm d}n= A m_0^{9/2} E^{-5/2} {\rm d}E ~~(\rho\gg\rho_{\rm H})~,
\label{edlher}
\end{equation}
where $A$ is a numerical coefficient independent of $m_0$ and of
$\rho$. Equation (\ref{edlher}) implies that the mean-square radius
$R$ of the sub-horizon loops is
\begin{equation}
R\sim m_0^{-3/2}E^{1/2}~.
\label{msr}
\end{equation}
Hence the large string loops are random walks of step approximately $
m_0^{-1}$.
Equations (\ref{edlher}) and (\ref{msr}) imply 
\begin{equation}
{\rm d}n= A' R^{-4} {\rm d}R ~~(\rho\gg\rho_{\rm H})~,
\label{sidclher}
\end{equation}
where $A'$ is a numerical constant.  Thus, at the high energy density
regime, the distribution of closed string loops is scale invariant,
since it does not depend on the cutoff parameter.  The total energy
density in sub-horizon string loops is independent of
$\rho$. Increasing the energy density $\rho$ of the system of strings,
the extra energy $E-E_{\rm H}$, where $E_{\rm H}=\rho_{\rm H}L^3$,
goes into the formation of super-horizon long strings, implying
\begin{equation}
\rho-\rho_{\rm l}={\rm const}  ~~(\rho\gg\rho_{\rm H})~,
\label{rhoinf}
\end{equation}
where $\rho_{\rm l}$ denotes the energy density in super-horizon 
loops (often called in the literature as {\sl infinite} strings). 

Clearly, the above analysis describes the behaviour of a system of
strings of low or high energy densities, while there is no analytic
description of the phase transition and of the intermediate densities
around the critical one, $\rho\sim\rho_{\rm H}$. An experimental
approach to the problem has been proposed in Ref.~\cite{smma1} and
later extended in Ref.~\cite{smed}.

The equilibrium properties of a system of cosmic strings have been
studied numerically in Ref.~\cite{smma1}. The strings are moving in a
three-dimensional flat space and the initial string states are chosen
to be a {\sl loop gas} consisting of the smallest two-point loops with
randomly assigned positions and velocities. This choice is made just
because it offers an easily adjustable string energy density. Clearly,
the equilibrium state is independent of the initial state. The
simulations revealed a distinct change of behaviour at a critical
energy density $\rho_{\rm H}=0.0172\pm 0.002$. For $\rho <\rho_{\rm
  H}$, there are no super-horizon strings, their energy density,
$\rho_{\rm l}$, vanishes. For $\rho >\rho_{\rm H}$, the energy density
in sub-horizon string loops is constant, equal to $\rho_{\rm H}$,
while the extra energy goes to the super-horizon string loops with
energy density $\rho_{\rm l}=\rho-\rho_{\rm H}$.  Thus,
Eqs.~(\ref{edlher}) and (\ref{rhoinf}) are valid for all
$\rho>\rho_{\rm H}$, although they were derived only in the limit
$\rho\gg\rho_{\rm H}$.  At the critical energy density,
$\rho=\rho_{\rm H}$, the system of strings is scale-invariant. At
bigger energy densities, $\rho>\rho_{\rm H}$, the energy distribution
of sub-horizon string loops at different values of $\rho$ were found
~\cite{smma1} to be identical within statistical errors, and
well-defined by a line ${\rm d}n/{\rm d}E\propto E^{-5/2}$. Thus, for
$\rho>\rho_{\rm H}$, the distribution of sub-horizon string loops is
still scale-invariant, but in addition the system includes
super-horizon string loops, which do not exhibit a scale-invariant
distribution.  The number distribution for super-horizon loops goes as
${\rm d}n/{\rm d}E\propto 1/E$, which means that the total number of
super-horizon string loops is roughly $\log(E-E_{\rm H})$. So,
typically the number of long strings grows very slowly with energy;
for $\rho>\rho_{\rm H}$ there are just a few super-horizon strings,
which take up most of the energy of the system.

The above numerical experiment has been extended~\cite{smed} for
strings moving in a higher dimensional box. The Hagedorn energy
density was found for strings moving in boxes of dimensionality
$d_{\rm B}=3,4,5$~\cite{smed}:
\begin{eqnarray}
\rho_{\rm H} &=&
\left\{ \begin{array}{lcl}0.172\pm 0.002  \ \ \mbox{for}\ \ & d_{\rm B}=3\\
0.062\pm 0.001  \ \ \mbox{for}\ \ & d_{\rm B}=4\\
0.031\pm 0.001  \ \ \mbox{for}\ \ & d_{\rm B}=5
\end{array}\right.
\end{eqnarray}
Moreover, the size distribution of sub-horizon string loops at the
high energy density regime was found to be independent of the
particular value of $\rho$ for a given dimensionality of the box
$d_{\rm B}$.  The size distribution of sub-horizon string loops was
found~\cite{smed} to be well defined by a line
\begin{equation}
\frac{dn}{dE}\sim E^{-(1+d_{\rm B}/2)}~,
\end{equation}
where the space dimensionality $d_{\rm B}$ was taken equal to 3, 4, or
5 . The statistical errors indicated a slope equal to $-(1+d_{\rm
B}/2)\pm 0.2$. Above the Hagedorn energy density the system is again
characterised by a scale-invariant distribution of sub-horizon string
loops and a number of super-horizon string loops with a distribution
which is not scale invariant.

\section{Cosmic String Gravity}

The gravitational properties of cosmic strings are very different than
those of non-relativistic linear mass distributions.  Straight cosmic
strings produce no gravitational force in the surrounding matter:
$\nabla ^2\Phi=0$, where $\Phi$ stands for the Newtonian
potential. Cosmic strings have relativistic motion, implying
that oscillating string loops can be strong emitters of gravitational
radiation. Since super-horizon cosmic strings have wiggles and
small-scale structure due to string intercommutations, they are also
sources of gravitational radiation~\cite{Sakellariadou:1991sd}.  A
gravitating string is described by a coupled system of Einstein, Higgs
and gauge field equations, for which no exact solution is known. We
thus usually make two simplifications: we consider the cosmic string
thickness to be much smaller than any other relevant dimension and the
cosmic string gravitational field to be sufficiently weak, so that
linearised Einstein equations can be used (for $G\mu\ll 1$).  

The geometry around a straight cosmic string is locally identical to
that of flat space-time, but this geometry is not globally Euclidean;
the azimuthal coordinate varies in the range $[0, 2\pi(1-4G\mu))$.
Hence, the effect of a cosmic string is to introduce an azimuthal {\sl
deficit angle} $\Delta$, whose magnitude is determined by the symmetry
breaking scale $T_{\rm c}$ leading to the cosmic string formation,
namely $\Delta =8\pi G\mu$. Thus, the string metric ${\rm d}s^2= {\rm
d}t^2-{\rm d}z^2-{\rm d}r^2-(1-8\pi G\mu)r^2{\rm d}\theta^2$ describes
a conical space leading to interesting observational effects on the
propagation of light (i.e., double images of light sources located
behind cosmic strings) and of particles (i.e., discontinuous Doppler
shift effects).  The centre-of-mass velocity $u$ of two particles,
moving towards a string at the same velocity $v$ reads
\begin{equation}
u=v\sin(\Delta/2)[1-v^2\cos^2(\Delta/2)]^{-1/2}\ .
\end{equation}
Considering that one of the particles carries a light source, while
the other one is an observer, one realises that the observer will
detect a discontinuous change in the frequency $\omega$ of light, given
by~\cite{ks}
\begin{equation}
{\delta\omega\over \omega}={v\over\sqrt{1-v^2}}\Delta\ .
\end{equation}
This discontinuous change in the frequency has its origin in the
Doppler shift: particles start moving towards each other, decreasing
their distance, once the line connecting the particles crosses the
string.

In the framework of gravitational instability, topological defects in
general and cosmic strings in particular, offered an alternative to
the inflationary paradigm for the origin of the initial fluctuations
leading to the observed large-scale structure and the measured
anisotropies of the CMB temperature anisotropies.  The angular power
spectrum of CMB is expressed in terms of the dimensionless
coefficients $C_\ell$, in the expansion of the angular correlation
function in terms of the Legendre polynomials $P_\ell$ reads 
\footnote{Equation (\ref{dtovertvs}) holds only if the initial state
  for cosmological perturbations of quantum-mechanical origin is the
  vacuum~\cite{Martin:1999fa}.}:
\begin{eqnarray}
&&\biggl \langle 0\biggl |\frac{\delta T}{T}({\bf n})\frac{\delta T}{
T}({\bf n}') \biggr |0\biggr\rangle \left|_{{~}_{\!\!({\bf n\cdot
n}'=\cos\vartheta)}}\right.\nonumber\\
&&\ \ \ \ \ \ \ \   = \frac{1}{4\pi}\sum_\ell(2\ell+1)C_\ell
P_\ell(\cos\vartheta) {\cal W}_\ell^2 \;,
\label{dtovertvs}
\end{eqnarray}
where ${\cal W}_\ell $ stands for the $\ell$-dependent window function
of the particular experiment.  Equation (\ref{dtovertvs}) compares
points in the sky separated by an angle $\vartheta$.  The value of
$C_\ell$ is determined by fluctuations on angular scales of the order
of $\pi/\ell$. The angular power spectrum of anisotropies observed
today is usually given by the power per logarithmic interval in
$\ell$, plotting $\ell(\ell+1)C_\ell$ versus $\ell$.

To find the power spectrum induced by topological defects, one has to
solve in Fourier space, for each given wave vector ${\bf k}$, a system
of linear perturbation equations with random sources:
\begin{equation}
{\cal D} X = {\cal S}\;,
\label{deq}
\end{equation}
where ${\cal D}$ denotes a time dependent linear differential
operator, $X$ is a vector which contains the various matter
perturbation variables, and ${\cal S}$ is the random source term,
consisting of linear combinations of the energy momentum tensor of the
defect.  For given initial conditions, Eq.~(\ref{deq}) can be solved
by means of a Green's function, ${\cal G}(\tau,\tau ')$.
To compute power spectra or, more generally, quadratic expectation
values of the form $\langle X_j(\tau_0,{\bf k})X_m^*(\tau_0,{\bf
k'})\rangle$, one has to calculate
\begin{eqnarray}
&&\langle X_j(\tau_0,{\bf k})X_l^\star(\tau_0,{\bf k}')\rangle 
= \int_{\tau_{in}}^{\tau_0}\! d\tau{\cal
G}_{jm}(\tau,{\bf k})\nonumber\\
&& \times \int_{\tau_{in}}^{\tau_0} \! d\tau '{\cal
G}^\star_{ln}(\tau ',{\bf k}') \langle{\cal S}_m(\tau,{\bf
k}){\cal S}_n^\star(\tau ',{\bf k}')\rangle\;.
\label{power}
\end{eqnarray}
To compute power spectra, one should know the unequal time two-point
correlators $\langle{\cal S}_m(\tau,{\bf k}){\cal S}_n^\star(\tau
',{\bf k}')\rangle$ in Fourier space, calculated by means of heavy
numerical simulations.

The first determinations of the CMB power spectrum from cosmic strings
were based on the assumption that cosmic strings are of infinitesimal
width. They were thus realised by either employing Nambu-Goto
simulations of connected string segments, or a model involving a
stochastic ensemble of unconnected segments.  However, a number of
questions there have been later raised regarding the accuracy of these
approaches. More recently, the CMB power spectrum contribution from
cosmic strings has been addressed~\cite{Bevis:2006mj} using
field-theoretic simulations of the Abelian Higgs model, the simplest
example of an underlying field theory with local U(1) strings.  All
approaches agree on the basic form of the cosmic string power
spectrum, namely a power spectrum with a roughly constant slope at low
multipoles, rising up to a single peak, and consequently decaying at
small scales. This result is common of all models, in which
fluctuations are generated continuously and evolve according to
inhomogeneous linear perturbation equations.

In topological defects models, fluctuations are constantly generated
by the non-linear defect evolution. This characteristic, combined with
the fact that the random initial conditions of the source term of a
given scale leak into other scales, destroy perfect coherence. The
incoherent aspect of active perturbations  affects the structure of
secondary oscillations, namely secondary oscillations may get washed
out. Thus, in topological defects models, incoherent fluctuations lead
to a single bump at smaller angular scales (larger $\ell$), than those
predicted within any inflationary scenario.

The cosmic string CMB power spectrum was found~\cite{Bevis:2006mj} to
have a broad peak at $\ell\approx 150-400$.  Decomposing the power
spectrum into scalar, vector and tensor modes, it was
shown~\cite{Bevis:2006mj} that the origin of this broad peak lies in
both the vector and scalar modes, which peak at $\ell\approx 180$ and
$\ell\approx 400$, respectively. This analysis
concluded~\cite{Bevis:2006mj} that the cosmic string power spectrum is
dominated by vector modes for all but the smallest scales.

The position and amplitude of the acoustic peaks, as found by the CMB
measurements, are in disagreement with the predictions of topological
defect  models. As a consequence, CMB measurements rule out pure
topological defect models in general, and cosmic strings in
particular, as the origin of initial density perturbations leading to
the observed structure formation.

Since cosmic strings are expected to be generically formed in the
context of SUSY GUTs, one should consider {\it mixed perturbation
  models} where the dominant r\^ole is played by the inflaton field
but cosmic strings have also a contribution, small but not negligible.
Restricting ourselves to the angular power spectrum, we can remain in
the linear regime. In this case,
\begin{equation}
C_\ell =   \alpha     C^{\scriptscriptstyle{\rm I}}_\ell
         + (1-\alpha) C^{\scriptscriptstyle{\rm S}}_\ell\;,
\label{cl}
\end{equation}
where $C^{\scriptscriptstyle{\rm I}}_\ell$ and $C^{\scriptscriptstyle
{\rm S}}_\ell$ denote the (COBE normalised) Legendre coefficients due
to adiabatic inflaton fluctuations and those stemming from the cosmic
string network, respectively. The coefficient $\alpha$ in
Eq.~(\ref{cl}) is a free parameter giving the relative amplitude for
the two contributions.  Comparing the $C_\ell$, calculated using
Eq.~(\ref{cl}) --  where $C^{\scriptscriptstyle{\rm I}}_\ell$ is taken
from a generic inflationary model and $C^{\scriptscriptstyle {\rm
S}}_\ell$ from numerical simulations of cosmic string networks -- 
with data obtained from the most recent CMB measurements, one gets
that a cosmic string contribution to the primordial fluctuations
higher than $14\%$ is excluded up to $95\%$ confidence
level~\cite{Bouchet:2000hd}.

Let us now return to F- and D-term hybrid inflation and investigate
the constraints on the free parameters of the model (namely masses and
couplings) so that the cosmic string contribution to the CMB data is
within the allowed limits imposed from recent CMB measurements.
Considering only large angular scales one can get the contributions to
the CMB temperature anisotropies analytically.  The quadrupole
anisotropy has one contribution coming from the inflaton field, and
one contribution coming from the cosmic string network.  Fixing the
number of e-foldings to 60, the inflaton and cosmic string
contribution to the CMB depend on the parameters of the model.  For
F-term inflation the cosmic string contribution to the CMB data is
consistent with CMB measurements provided~\cite{Rocher:2004et}
\begin{equation}
M\lsim 2\times 10^{15} {\rm GeV} ~~\Leftrightarrow ~~\kappa \lsim
7\times10^{-7}~.
\end{equation}
The superpotential coupling $\kappa$ is also subject to the gravitino
constraint which imposes an upper limit to the reheating temperature,
to avoid gravitino overproduction. Within the framework of SUSY GUTs
and assuming a see-saw mechanism to give rise to massive neutrinos,
the inflaton field decays during reheating into pairs of right-handed
neutrinos.  This constraint on the reheating temperature can be
converted to a constraint on the parameter $\kappa$. The gravitino
constraint on $\kappa$ reads~\cite{Rocher:2004et} $\kappa \lsim 8\times
10^{-3}$, which is rather weaker.

The tuning of $\kappa$ can be softened if one allows for the curvaton
mechanism.  The curvaton is a scalar field that is sub-dominant during
the inflationary era as well as at the beginning of the radiation
dominated era following inflation. In the context of supersymmetric
theories such scalar fields are expected to exist, and in addition, if
embedded strings accompany the formation of cosmic strings, they may
offer a natural curvaton candidate, provided the decay product of
embedded strings gives rise to a scalar field before the onset of
inflation.  Assuming the existence of a curvaton field there is an
additional contribution to the temperature anisotropies and the CMB
measurements impose~\cite{Rocher:2004et} the following limit on the
initial value of the curvaton field
$${\cal\psi}_{\rm init} \lsim 5\times 10^{13}\,\left(
\frac{\kappa}{10^{-2}}\right){\rm GeV}~~\mbox{for}~~ \kappa\in
     [10^{-6},~1]~.$$

\begin{figure}[hhh]
\begin{center}
\includegraphics[scale=.4]{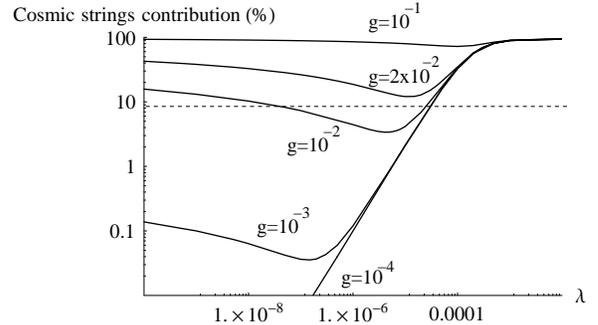}
\caption{For D-term inflation in minimal SUGRA, cosmic string
  contribution to CMB quadrupole anisotropies as a function of the
  superpotential coupling constant $\lambda$, for various values of
  the gauge coupling $g$. Figure taken from
  Ref.~\cite{Rocher:2004et}.}\label{Fig:Dtermstandard}
\end{center}
\end{figure}

D-term inflation can also be compatible with CMB measurements, provide
we tune its free parameters.  In the case of minimal SUGRA,
consistency between CMB measurements and theoretical predictions
impose~\cite{Rocher:2004et,Rocher:2004my} that $g\lsim 2\times
10^{-2}~ ~ \mbox{and}~ ~ \lambda\lsim 3\times 10^{-5}$, which can be
expressed as a single constraint on the Fayet-Iliopoulos term $\xi$,
namely $\sqrt\xi \lsim 2\times 10^{15}~{\rm GeV}.$ The results are
illustrated in Fig.\ref{Fig:Dtermstandard}.
The fine tuning on
the couplings can be softened if one invokes the curvaton mechanism
and constrains the initial value of the curvaton field to
be~\cite{Rocher:2004my}
$$\psi_{\rm init}\lsim 3\times
10^{14}\left(\frac{g}{10^{-2}}\right){\rm
  GeV}~~\mbox{for}~~\lambda\in[10^{-1},10^{-4}]~.$$ For D-term
inflation based on K\"ahler geometry with shift symmetry, the cosmic
string contribution to the CMB anisotropies is dominant, in
contradiction with the CMB measurements, unless the superpotential
coupling is~\cite{Rocher:2004my} $\lambda\lsim 3\times 10^{-5}.$
Finally, in the case of D-term inflation based on a K\"ahler potential
with non-renormalisable terms, the contribution of cosmic strings
dominates if the superpotential coupling $\lambda$ is close to unity.
The constraints on $\lambda$ read~\cite{Rocher:2004my}
$$(0.1-5)\times 10^{-8} 
 \leq \lambda \leq
(2-5)\times 10^{-5}$$ 
or, equivalently
$$\sqrt{\xi}\leq 2\times 10^{15} \;\mathrm{GeV}~,
$$ implying $ G\mu \leq 8.4\times 10^{-7}$.  Thus, higher order
K\"ahler potentials do not suppress cosmic string contribution.

Apart the temperature power spectrum, important constraints on cosmic
string scenarios might also arise in the future from measurements of
the polarisation of the CMB photons. More precisely, the
B-polarisation spectrum offers an interesting window on cosmic
strings~\cite{Bevis:2007qz} since inflation has only a weak
contribution. Scalar modes may contribute to the B-mode only via the
gravitational lensing of the E-mode signal, with a second inflationary
contribution coming from the sub-dominant tensor modes.

Cosmic strings can also become apparent through their contribution in
the small-angle CMB temperature anisotropies.  More precisely, at high
multipoles $\ell$ (small angular resolution), the mean angular power
spectrum of string-induced CMB temperature anisotropies can be
described~\cite{Fraisse:2007nu} by $\ell^{-\alpha}$, with $\alpha \sim
0.889$.  Thus, a non-vanishing cosmic string contribution to the
overall CMB temperature anisotropies may dominate at high multipoles
$\ell$ (small angular scales).  In an arc-minute resolution
experiment, strings may be observable~\cite{Fraisse:2007nu} for $G\mu$
down to $2\times 10^{-7}$.

Cosmic strings should also induce deviations from Gaussianity. On
large angular scales such deviations are washed out due to the low
string contribution, however on small angular scales, optimal
non-Gaussian string-devoted statistical estimators may impose severe
constraints on a possible cosmic string contribution to the CMB
temperature anisotropies.

Finally, let me emphasise  that one should keep in mind that all
string-induced CMB temperature anisotropies were performed for Abelian
strings in the zero thickness limit with reconnection probability
equal to unity and winding number equal to one. Even though in any
model where fluctuations are constantly induced by sources ({\sl
seeds}) having a non-linear evolution, the perfect coherence which
characterises the inflationary induced spectrum of perturbations gets
destroyed~\cite{ruth-mairi},
there is still no reason to expect that quantitatively the results
found for conventional cosmic string models will hold in more general
cases.

\section{Superconducting Cosmic Strings}
Before finishing this brief review on cosmic strings, let me mention
the case of superconducting strings~\cite{witten}, in the sense that
in a large class of high energy physics theories, strings have similar
electromagnetic properties as those of superconducting wires. Such
objects carry large electric currents and hence their interaction with
the cosmic plasma can lead to a variety of distinct astrophysical
effects.

Cosmic strings are characterised as superconductors if electromagnetic
gauge invariance is broken inside the strings, a situation which can
occur for instance when a charged scalar field develops a non-zero
expectation value in the vicinity of the string core \footnote{Note
that also a vector field, whose flux is trapped inside a non-Abelian
string can lead to a superconducting string.}.  Superconducting
strings appear also in models with fermions, which acquire masses
through a Yukawa coupling to the Higgs field of the strings. Thus,
depending on the considered model we can have bosonic or fermionic
string superconductivity. In the first case, bosons can condensate and
acquire a non-vanishing phase gradient, while in the second one,
fermions may propagate in the form of zero modes along the string.

Applying an electric field on a superconducting string, the string
will develop a growing electric current according to
\begin{equation}
\frac{{\rm d}J}{{\rm d} t}\sim\frac{c e^2}{\hbar} E\ ,
\end{equation}
where $E$ stands for the field component along the string and $e$
denotes the elementary charge.

In the case of fermionic superconductivity, fermions are massless
inside the string, whereas they have a finite mass, $m$, outside the
string \footnote{The fermion mass is model-dependent, but it is
bounded from above by the symmetry breaking scale of the string.}.
Under the effect of an electric field, a current $J$ results, growing
in time until it reaches a critical value $J_{\rm c}\sim emc^2/\hbar$,
when the particles inside the string, moving at relativistic speeds,
have sufficient energy to leave the string. Thus, when the string
current reaches its critical value, $J_{\rm c}$, particles get
produced at a rate
\begin{equation}
\dot n\sim eE/\hbar\ ,
\end{equation}
where $n$ stands for the number of fermions per unit length.  Note
that $J_{\rm c}$, even though model-dependent, it does not exceed a
maximum value, given by $J_{\rm max}\sim e(\mu c^3/\hbar)^{1/2}.$
Depending on their energy scale, superconducting strings may carry
huge currents.  In the case of bosonic superconductivity, the
model-dependent critical current $J_{\rm c}$ --- determined by the energy
scale at which scale invariance is broken --- is again bounded by $J_{\rm
max}$, defined as above.

Superconducting strings can also develop growing currents in magnetic
fields, according to
\begin{equation}
\frac{{\rm d}J}{{\rm d} t}\sim\frac{e^2}{\hbar} vB\ ,
\end{equation}
where $v$ is the speed of the moving string segment in a magnetic
field $B$. For a string loop carrying sufficiently large currents,
electromagnetic radiation can overtake gravitational radiation, 
becoming the dominant energy loss mechanism.

Superconducting string loops may be problematic in cosmology.  For a
current-carrying loop, the energy per unit length is not equal to the
tension, their difference equals the string current. Such a loop can
rotate and the resulting centrifugal force --- which can be expected
to be very much stronger than the inefficient magnetic {\sl spring}
repulsion effect --- may balance the tension. When a rotating string
loop reaches an equilibrium state --- defined by the balance between
the string tension and the centrifugal force --- it is called a {\sl
  vorton}~\cite{vortons}.  Vortons can be formed at, or soon after,
the phase transition followed by SSB leading to the string formation,
and they possess a net charge as well as a current.  If vortons are
stable (certainly a model-dependent issue~\cite{cm}), they will scale as matter
in the Universe, dominating over its energy density.  In this
sense, ortons may constrain models for superconducting strings.

\section{Cosmic Superstrings}
The recent interplay between superstring theory and cosmology has led
to the notion of cosmic superstrings~\cite{Polchinski:2004ia},
providing the missing link between superstrings and their classical
analogues.

The possible astrophysical r\^ole of superstrings has been advocated
already more than twenty years ago. More precisely, it has been
proposed~\cite{Witten1985}, that superstrings of the O(32) and
E$_8\times$E$_8$ string theories are likely to generate string-like
stable vortex lines and flux tubes.  However, in the context of
perturbative string theory, the high tension (close to the Planck
scale) of fundamental strings ruled them out~\cite{Witten1985} as
potential cosmic string candidates.  Luckily, this picture has changed
in the framework of brane-world cosmology, which offers an elegant
realisation of nature within string theory. Within the brane-world
picture, all standard model particles are open string modes. Each end
of an open string lies on a brane, implying that all standard model
particles are stuck on a stack of D$p$-branes, while the remaining
$p-3$ of the dimensions are wrapping some cycles in the bulk. Closed
string modes (e.g., dilaton, graviton) live in the high-dimensional
bulk.  Brane interactions lead to unwinding and thus evaporation of
higher dimensional D$p$-branes. We are eventually left with D3-branes
--- one of which could indeed play the r\^ole of our
Universe~\cite{Durrer:2005nz} --- embedded in a
(9+1)-dimensional bulk and cosmic superstrings (one-dimensional
D-branes, called D-strings, and Fundamental strings, called F-strings).

Brane annihilations provide a natural mechanism for ending
inflation. To illustrate the formation of cosmic superstrings at the
end of brane inflation, let us consider a D$p$-${\bar D}p$
brane-anti-brane pair annihilation to form a D$(p-2)$ brane.  Each
{\sl parent} brane has a U(1) gauge symmetry and the gauge group of
the pair is U(1)$\times$U(1). The {\sl daughter} brane possesses a
U(1) gauge group, which is a linear combination, U(1)$_-$, of the
original two U(1)'s. The branes move towards each other and as their
inter-brane separation decreases below a critical value, the tachyon
field, which is an open string mode stretched between the two branes,
develops an instability. The tachyon couples to the combination
U(1)$_-$. The rolling of the tachyon field leads to the decay of the
{\sl parent} branes. Tachyon rolling leads to spontaneously symmetry
breaking, which supports defects with even co-dimension.  So, brane
annihilation leads to vortices, D-strings; they are cosmologically
produced via the Kibble mechanism.  The other linear combination,
U(1)$_+$, disappears since only one brane remains after the brane
collision.  The U(1)$_+$ combination is thought to disappear by having
its fluxes confined by fundamental closed strings.  Such strings are
of cosmological size and they could play the r\^ole of cosmic
strings~\cite{Sarangi:2002yt}; they are referred to in the literature
as cosmic superstrings~\cite{Polchinski:2004hb}.

\section{Differences between Cosmic Strings and Cosmic Superstrings}

Cosmic superstrings~\cite{Polchinski:2004ia}, even though
cosmologically extended, are quantum objects, in contrast to solitonic
cosmic strings which are classical objects. Hence, one expects a
number of differences to arise as regarding the properties of the two
classes of objects.  As I have earlier discussed, the probability
that a pair of cosmic strings will reconnect, after having intersect, 
equals unity.  The reconnection probability for cosmic
superstrings is however smaller (often much smaller) than unity. The
corresponding intercommutation probabilities are calculated in string
perturbation theory. The result depends on the type of strings and on
the details of compactification. For fundamental strings, reconnection
is a quantum process and takes place with a probability of order
$g_{\rm s}^2$ (where $g_{\rm s}$ denotes the string tension).  It can
thus be much less than one, leading to an increased density of
strings~\cite{Sakellariadou:2004wq}, implying an enhancement of
various observational signatures.  The reconnection probability is a
function of the relative angle and velocity during the collision.  One
may think that strings can miss each other, as a result of their
motion in the compact space. Depending on the supersymmetric
compactification, strings can wander over the compact dimensions, thus
missing each other, effectively decreasing their reconnection
probability.  However, in realistic compactification schemes, strings
are always confined by a potential in the compact
dimensions~\cite{Jackson:2004zg}. The value of $g_{\rm s}$ and the
scale of the confining potential will determine the reconnection
probability. Even though these are not known, for a large number of
models it was found~\cite{Jackson:2004zg} that the reconnection
probability for F-F collisions lies in the range between $10^{-3}$ and
1. The case of D-D collisions is more complicated; for the same models
the reconnection probability is anything between 0.1 to 1. Finally,
the reconnection probability for F-D collisions can vary from 0 to 1.

Brane collisions lead not only to the formation of F- and D-strings,
they also produce bound states, $(p,q)$-strings, which are composites
of $p$ F-strings and $q$ D-strings~\cite{Copeland:2003bj}.  The
presence of stable bound states implies the existence of junctions,
where two different types of string meet at a point and form a bound
state leading away from that point.  Thus, when cosmic superstrings of
different types collide, they can not intercommute, instead they
exchange partners and form a junction at which three string
segments meet. This is just a consequence of charge conservation at the
junction of colliding $(p, q)$-strings.  For $p = np'$ and $q = nq'$,
the $(p, q)$ string is neutrally stable to splitting into $n$ bound
$(p', q')$ strings.  The angles at which strings pointing into a
vertex meet, is fixed by the requirement that there be no force on the
vertex.  In general, a $(p, q)$ and a $(p', q')$ string will form a
trilinear vertex with a $(p+p', q+q')$ or a $(p-p', q-q')$
string. This leads certainly to the crucial question of whether a
cosmic superstring network will reach scaling, or whether it freezes
leading to predictions inconsistent with our observed Universe
universe.

The tension of solitonic strings is set from the energy scale of the
phase transition, followed by a spontaneously broken symmetry, which
left behind these defects as false vacuum remnants.  Cosmic
superstrings however span a whole range of tensions, set from the particular
brane inflation model.  The tension of F-strings in 10 dimensions is
$\mu_{\rm F}=1/(2\pi\alpha')$, and the tension of D-strings is
  $\mu_{\rm D}=1/(2\pi\alpha' g_{\rm s})$, where $g_{\rm s}$ stands
  for the string coupling. In 10 flat dimensions, supersymmetry
  dictates that the tension of the $(p,q)$ bound states reads
\begin{equation}
\mu_{(p,q)}=\mu_{\rm F}\sqrt{p^2 + q^2/g_{\rm s}^2}~.
\label{squarelaw-tension}
\end{equation}
Individually, the F- and D-strings are ${1\over 2}$-BPS
(Bogomol'nyi-Prasad-Sommerfield) objects, which however break a
different half of the supersymmetry each. Equation
(\ref{squarelaw-tension}) represents the BPS bound for an object
carrying the charges of $p$ F-strings and $q$ D-strings.  Note that
the BPS bound is saturated by the F-strings, $(p,q)=(1,0)$, and the
D-strings, $(p,q)=(0,1)$.  Let me also make the remark that the string
tension for strings at the bottom of a throat is different from the
(simple) expression given in Eq.~(\ref{squarelaw-tension}) and it
depends on the choice of flux compactification.

Consider an F- and a D-string, both lying along the same 
axis. The total tension of this configuration is $(g_{\rm
s}^{-1}+1)/(2\pi\alpha')$, which exceeds the BPS bound; thus the
configuration is not supersymmetric. It can however lower its energy,
if the F-string breaks, its ends being attached to the D-string. Since
the end points can then move off at infinity, leaving only the
D-string behind, a flux will run between the end points of the
F-string. Now the tension reads $(g_{\rm s}^{-1}+{\cal{O}}(g_{\rm
s}))/(2\pi\alpha')$, and hence the final state represents a D-string
with a flux, which is a supersymmetric state.

\section{Cosmic Superstring Evolution}

The evolution of cosmic superstring networks, is a complicated issue,
which has been addressed by
numerical~\cite{Sakellariadou:2004wq,Avgoustidis:2004zt,Rajantie:2007hp,Sakellariadou:2008ay},
as well as analytical~\cite{Copeland:2006eh} approaches.

The first numerical attempt~\cite{Sakellariadou:2004wq}, studying
independent stochastic networks of D- and F-strings in a flat
space-time, has shown that the characteristic length scale $\xi$,
giving the typical distance between the nearest string segments and
the typical curvature of strings, grows linearly with time
\begin{equation}
 \xi(t)\propto \zeta t ~;
\end{equation}
the slope $\zeta$ depends on the reconnection probability ${\cal P}$,
and on the energy of the smallest allowed loops (i.e., the energy
cutoff).  For reconnection (or intercommuting) probability in the
range $10^{-3}\lsim {\cal P} \lsim 0.3$, it was
shown~\cite{Sakellariadou:2004wq} that
\begin{equation}
\zeta \propto \sqrt{\cal P} \Rightarrow \xi(t)\propto \sqrt{\cal P} t~.
\label{law}
\end{equation}

One can find in the literature (e.g., Ref.~[\cite{Sarangi:2002yt}c])
statements claiming that $\xi(t)$ should be instead proportional to
${\cal P} t$.  If this were correct, then the energy density of cosmic
superstrings, of given tension, could be considerably higher than that
of their field theory analogues.  However, the authors of
Ref.~\cite{Sarangi:2002yt} have missed out in their analysis that
intersections between two long strings is not the most efficient
mechanism for energy loss of the string network.  The findings of
Ref.~\cite{Sakellariadou:2004wq} cleared the misconception about the
behaviour of the scale $\xi$, and shown that the cosmic superstring
energy density may be higher than in the field theory case, but at most
only by one order of magnitude \footnote{A discussion and explanation
of this misconception can be found in Ref.~[\cite{vs}c].}.

As I have already discussed, in a realistic case $(p,q)$
strings come in a large number of different types, while a $(p,q)$
string can decay to a loop only if it self-intersects of collide with
another $(p,q)$ or $(-p,-q)$ string. A collision between $(p,q)$ and
$(p',q')$ strings will lead to a new $(p\pm p',q\pm q')$ string,
provided the end points of the initial two strings are not attached to
other three-string vertices, thus they are not a part of a web.  If
the collision between two strings can lead to the formation of one new
string, on a timescale much shorter than the typical collision
timescale, then the creation of a web may be avoided, and the
resulting network is composed by strings which are on the average
non-intersecting. Then one can imagine the following configuration: A
string network, composed by different types of $(p,q)$ strings
undergoes collisions and self-intersections. Energy considerations
imply the production of lighter daughter strings, leading eventually
to one of the following strings: $(\pm 1,0), (0,\pm 1), \pm(1,1),
\pm(1,-1)$. These ones may then self-intersect, form loops and scale
individually. Provided the relative contribution of each of these
strings to the energy density of the Universe is small enough, the
Universe will not be overclosed.

Let us now study the dynamics of a three-string junction in a simple
model.  The solutions of the BPS saturated formula
\begin{equation}
\mu_{(p,q)}=\sqrt{[p\mu_{(1,0)}]^2+[q\mu_{(0,1)}]^2}~,
\end{equation}
read
\begin{equation}
\mu_{(p,q)}\sin\alpha=q\mu_{(0,1)}\ \ ;
  \ \ \mu_{(p,q)}\cos\alpha=p\mu_{(1,0)}\ ,
\end{equation}
where $\tan\alpha=q/(pg_{\rm s})$.  The balance conditions for three
strings imply that when an F-string ends on a D-string, it causes it
to bend at an angle set by the string coupling; on the other side of
the junction there is a $(1,1)$ string.  Consider a junction of three
strings, with coordinates ${\bf x}(\sigma,t)$ , tension $\mu$ and
parameter lengths $L_1(t), L_2(t), L_3(t)$,which are joined at a
junction and whose other end terminate on parallel branes. The action
for this configuration reads \footnote{Note that in principle, cosmic
superstring dynamics ought to be studied using the Dirac-Born-Infeld
action, the low-energy effective action for many varieties of strings
arising in the context of string theory.}~\cite{Copeland:2006eh}
\begin{eqnarray}
\label{3stringjunction}
S=-\sum_{\alpha=1}^3 \mu_\alpha\int{\rm d}t\int_0^{L_\alpha(t)}{\rm
  d}\sigma \sqrt{-\gamma^{(\alpha)}}\nonumber\\ \ \ +
  \sum_{\alpha=1}^3\int{\rm d}t l_{\alpha}\cdot({\bf
    x}(t,L_\alpha(t)))-{\bf x}_{\rm junc}(t))~,
\end{eqnarray}
where the first part stands for the Nambu-Goto terms for the three
strings, and $l_{\alpha}$ denote the Lagrange multipliers to describe
the junction, located at position ${\bf x}_{\rm junc}$. From
Eq.~(\ref{3stringjunction}) one can derive the equations of motion as
well as the energy conservation.  One can easily check that
\begin{eqnarray}
\label{motion3stringjunction}
{\mu_1 (1-\dot L_1)\over
  \mu_1+\mu_2+\mu_3}=~~~~~~~~~~~~~~~~~~~~~~~~~~~~~~~~~~
\nonumber\\ {M_1(1-c_{23})\over
  M_1(1-c_{23})+M_2(1-c_{13})+M_3(1-c_{12})}~,
\nonumber\\ 
\nonumber\\ \mu_1\dot L_1+\mu_2\dot L_2+\mu_3\dot L_3=0~,~~~~~~~
\end{eqnarray}
and cyclic permutations. Note that $M_1=\mu_1^2-(\mu_2-\mu_3)^2$ (and
cyclic permutations giving $M_2, M_3$) and $c_{ij}={\bf
  a}_i'(t-L_i(t))\cdot {\bf a}_j'(t-L_j(t))$.  Equation
(\ref{motion3stringjunction}) implies that the rate of creation of new
string must balance the disappearance of old one. Thus, for an
F-string with $\mu_1=1$, and a D-string with $\mu_2=1/g_{\rm s}$, the
FD-bound state has tension $\mu_3=\sqrt{1+1/g_{\rm s}^2}=1/g_{\rm
  s}+g_{\rm s}/2+{\cal{O}}(g_{\rm s}^2)$. Since the angle $\alpha$
goes to $\pi/2$ in the limit of zero string coupling, we conclude that
in the small $g_{\rm s}$-limit, the length of the F-string remains
constant, while the length of the D-string decreases and the length of
the FD-bound state increases. This result has been recently confirmed
from numerical experiments~\cite{Sakellariadou:2008ay}.

To shed some light on the evolution of cosmic superstring networks, a
number of numerical experiments have been conducted, each of them at a
different level of approximation. One should keep in mind that the
initial configuration depends on the particular brane inflation
scenario, while a realistic network should contain strings with
junctions and allow for a spectrum of possible tensions.  

\begin{figure}[htbp] 
\begin{center} 
\includegraphics[width=0.205\textwidth,angle=0]{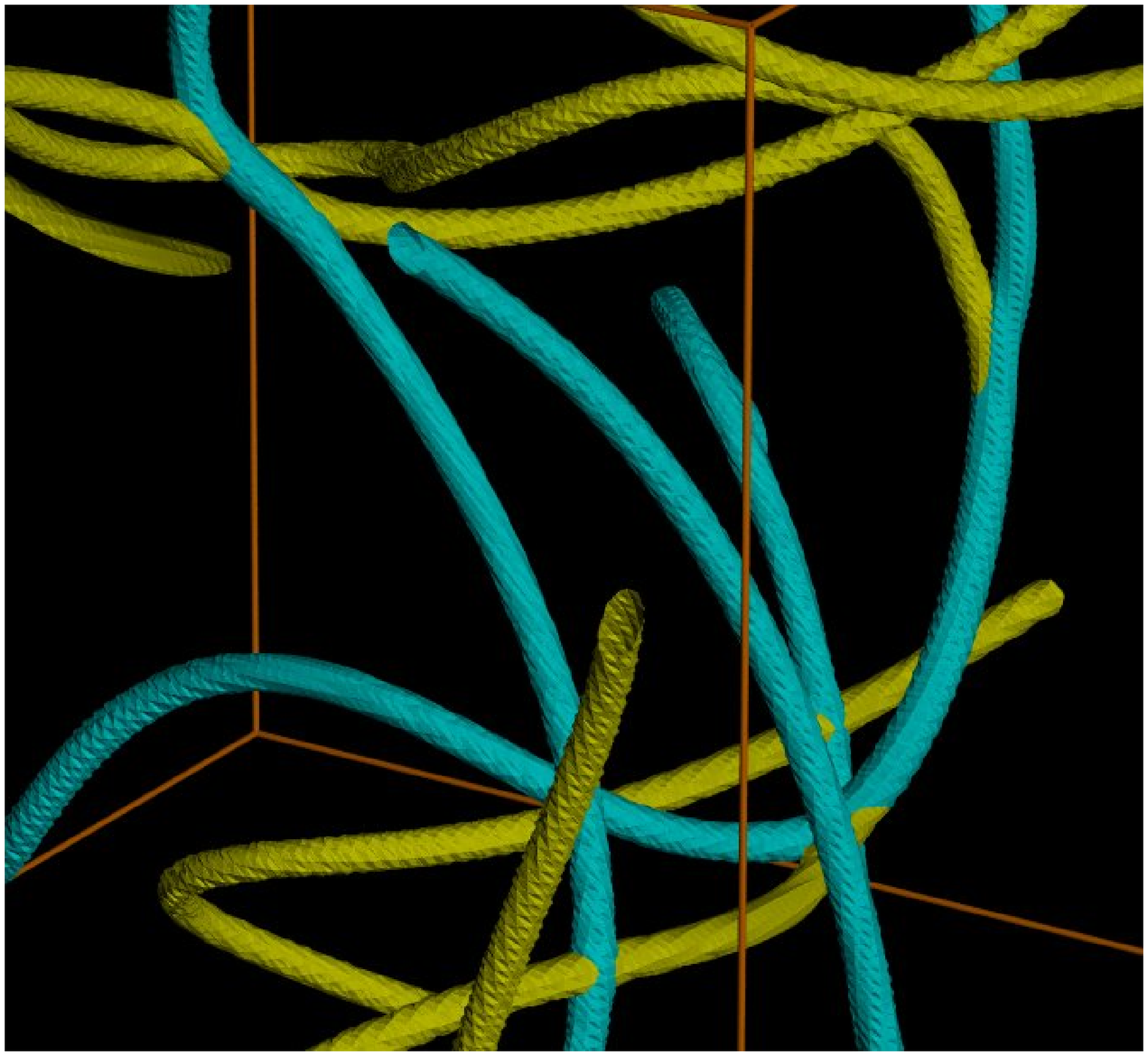} 
\includegraphics[width=0.23\textwidth,angle=0] {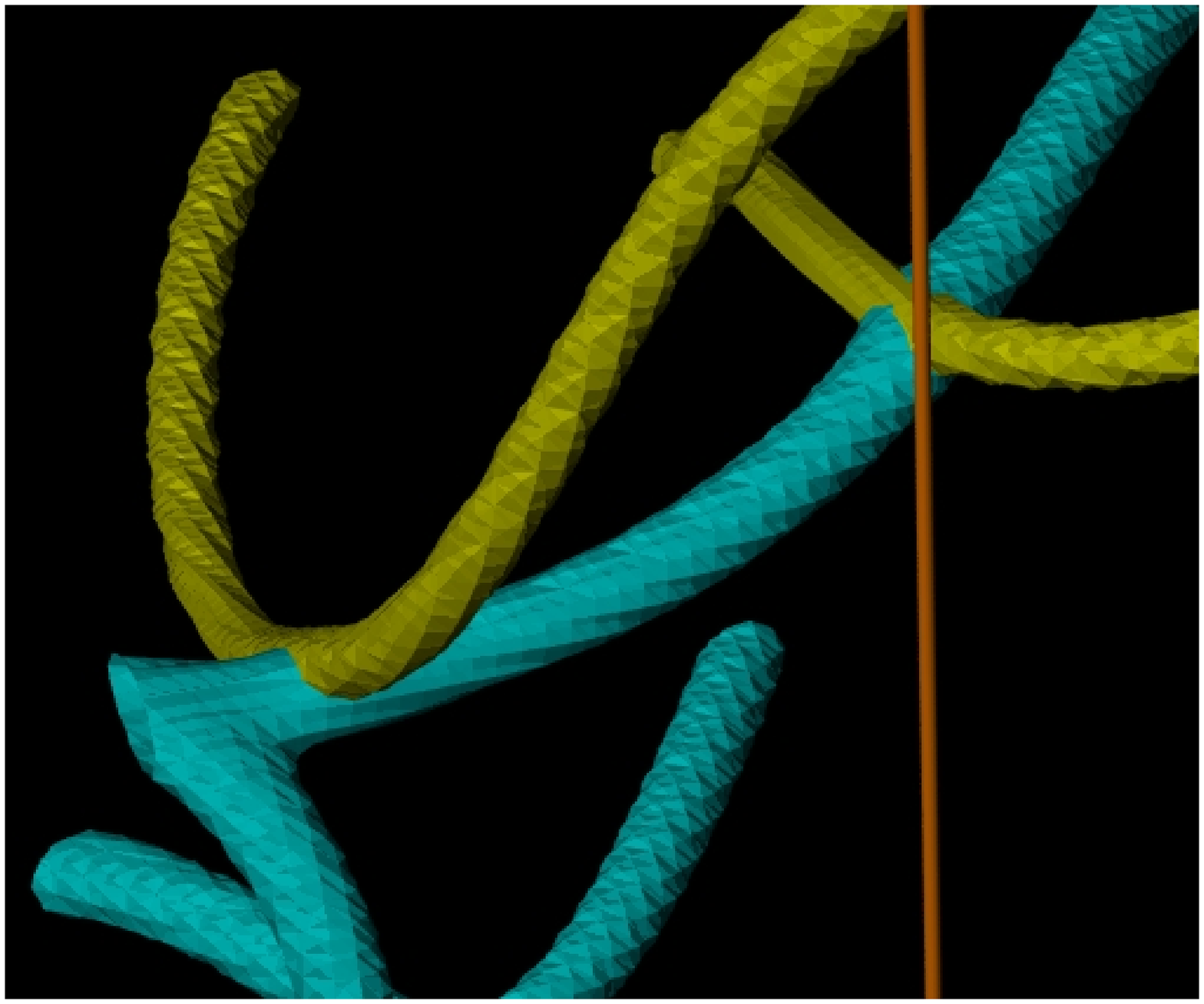} 
\end{center} 
\caption{Left: Bound states for local-local $\left(p,q\right)$ 
  strings. Right: Bound states for local-global $\left(p,q\right)$  
  strings. Figure taken from Ref.~\cite{Rajantie:2007hp}. 
  \label{bound_states_pictures}} 
\end{figure}

I will briefly describe the approach and findings of one of these
numerical approaches~\cite{Rajantie:2007hp}, which I consider more
realistic than others.  The aim of that study was to build a simple
field theory model of $(p,q)$ bound states, in analogy with the
Abelian Higgs model used to investigate the properties of solitonic
cosmic string networks, and to study the overall characteristics of
the network using lattice simulations.  Two models were investigated,
one in which both species of string have only short-range interactions
and another one in which one species of string features long-range
interactions.  We thus modelled the network with no long-range
interactions using two sets of fields, complex scalars coupled to
gauge fields, with a potential chosen such that the two types of
strings will form bound states ({\sl see},
Fig.~\ref{bound_states_pictures}). In this way junctions of 3 strings
with different tension were successfully modelled.  In order to
introduce long-range interactions we considered a network in which one
of the scalars forms global strings. This is important if the strings
are of a non-BPS species. For example, for cosmic superstrings at the
bottom of a Klebanov-Strassler throat the F-string is not BPS while
the D-string is.

More precisely, the $(p,q)$ string network was
modelled~\cite{Rajantie:2007hp} using two sets of Abelian Higgs
fields, $\phi, \chi$.  In the case that both species of cosmic strings
are BPS, the model is described by the action~\cite{Rajantie:2007hp}:
 \begin{eqnarray} 
\label{equ:action}
 \mathcal{S} &=& \int {\rm d}^{3}x{\rm d}t \biggl[\biggr.
   -\frac{1}{4}F^{2} -\frac{1}{2}\left(D_\mu\phi\right)
   \left(D^{\mu}\phi\right)^{*}\nonumber\\
  && ~~~~~~~~~~~~~-\frac{\lambda_{1}}{4}\left(\phi\phi^{*}-\eta_{1}^{2}\right)^{2}
   \nonumber \\ &&\  \ \ \ \ \ \ \ \ \ \ \ \ \ -\frac{1}{4}H^{2}
   -\frac{1}{2} \left(D_\mu\chi\right) \left(D^{\mu}\chi\right)^{*}\nonumber\\
   &&~~~~~~~~~~~~~-\frac{\lambda_{2}}{4}\phi\phi^{*}
   \left(\chi\chi^{*}-\eta_{2}^{2}\right)^{2} \biggl.\biggr]~,
\end{eqnarray} 
where the covariant derivative $D_\mu$ is defined by
\begin{eqnarray}
D_\mu\phi&=&\partial_\mu\phi-ie_1A_\mu\phi~,\nonumber\\
D_\mu\chi&=&\partial_\mu\chi-ie_2C_\mu\chi~.
\end{eqnarray}
For clarity, we label the $\phi$ field as ``Higgs'' and the $\chi$
field as ``axion'', even though both fields are Higgs-like.  The
scalars are coupled to the U(1) gauge fields $A_\mu$ and $C_\mu$, with
coupling constants $e_1$ and $e_2$ and field strength tensors
$F_{\mu\nu}=\partial_\mu A_\nu -\partial_\nu A_\mu$ and
$H_{\mu\nu}=\partial_\mu C_\nu -\partial_\nu C_\mu$, respectively.
The scalar potentials are parametrised by the positive constants
$\lambda_1, \eta_1$ and $\lambda_2, \eta_2$, respectively.
In the case that one species of string is non-BPS, we remove the
second gauge field by setting $e_2=0$. In this way, this species of
string is represented by the topological defect of a complex scalar
field with a global U(1) symmetry. Note that such defects are
characterised by the existence of long-range interactions --- as
opposed to local strings in which all energy density is confined
within the string, so that local strings have only gravitational
interactions --- implying different consequences for the evolution of
the network.

\begin{figure}[htbp] 
\begin{center} 
\includegraphics[width=0.23\textwidth,angle=0]{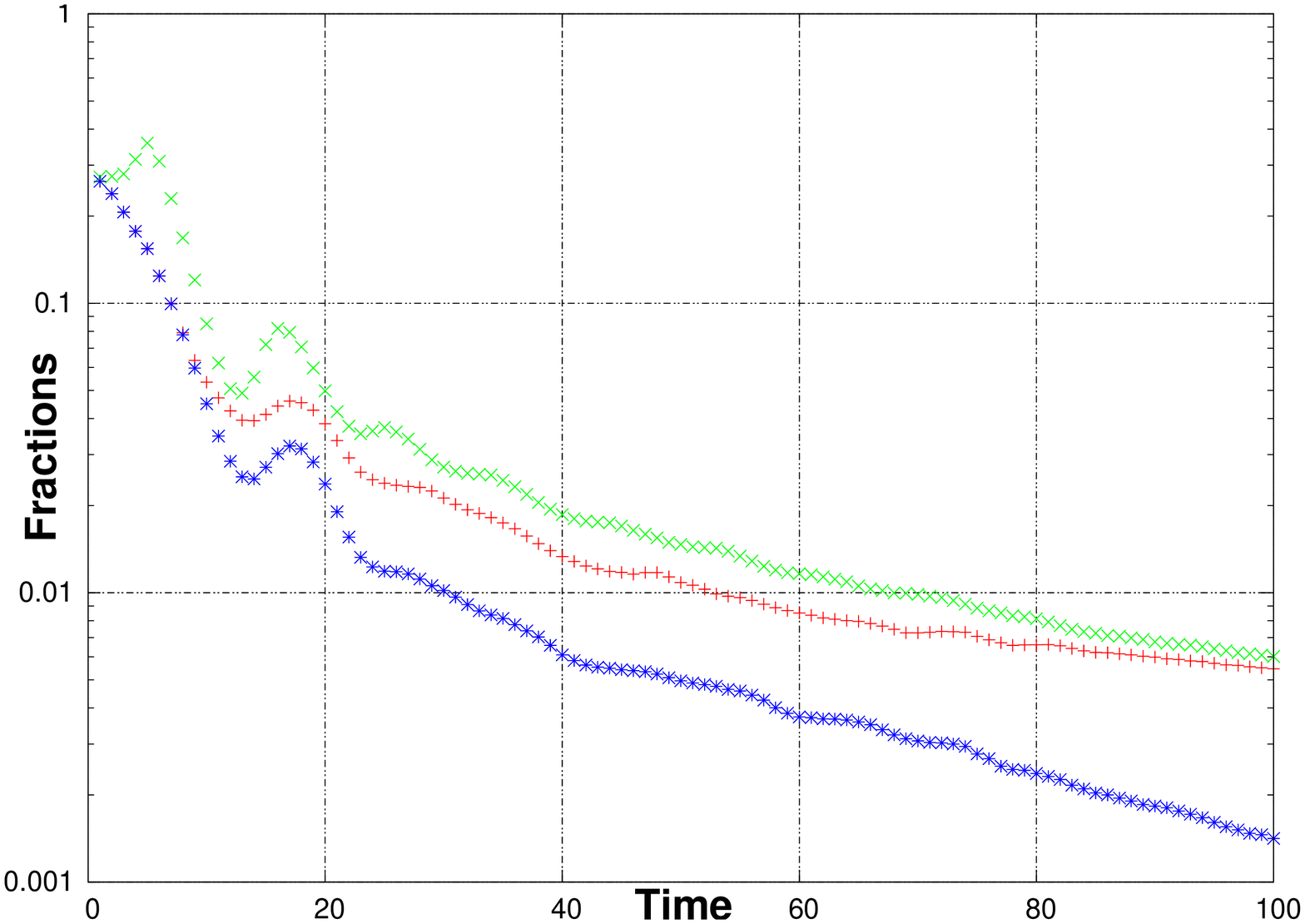}
\includegraphics[width=0.23\textwidth,angle=0]{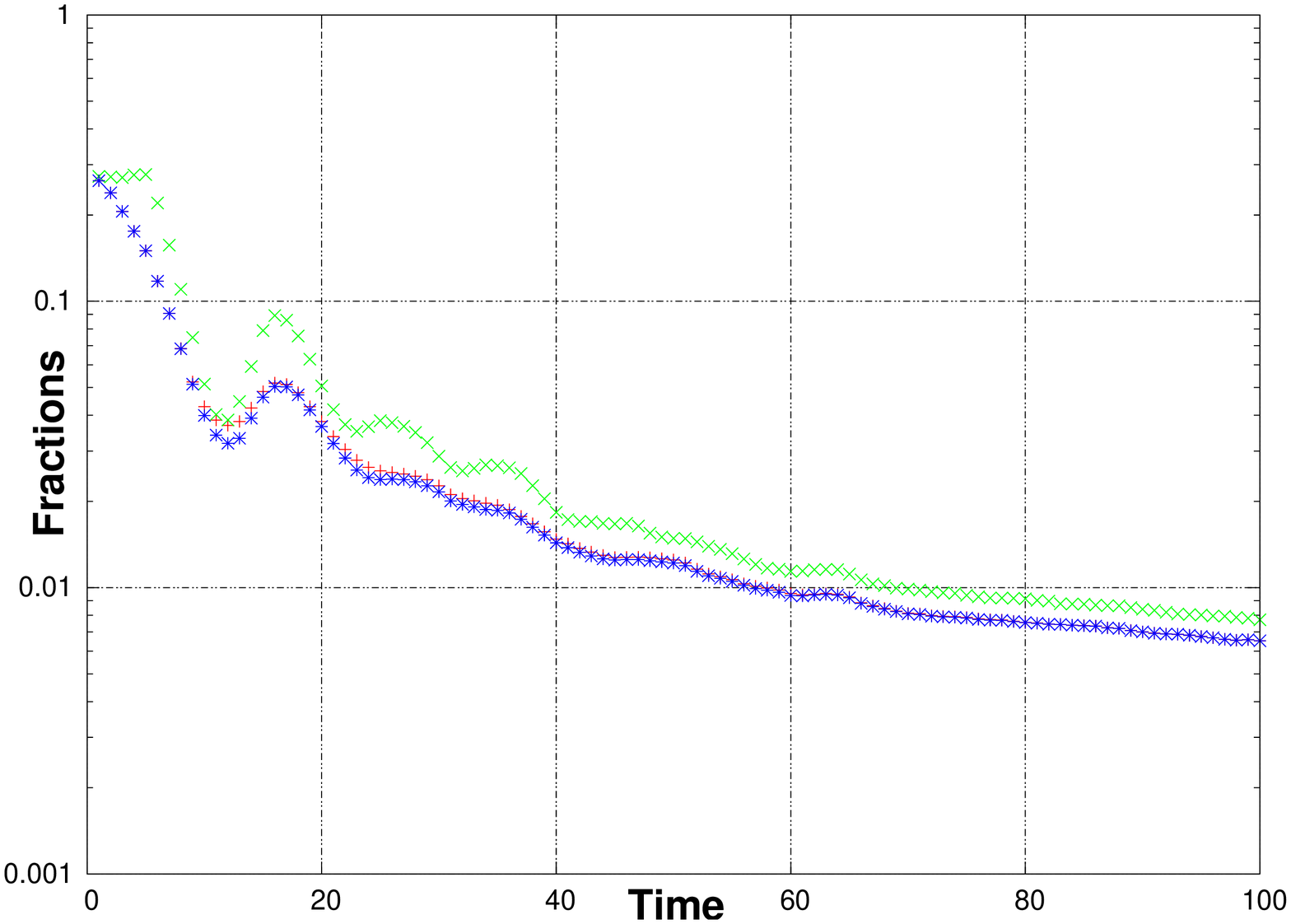}
\end{center} 
 
\begin{center} 
\includegraphics[width=0.23\textwidth,angle=0]{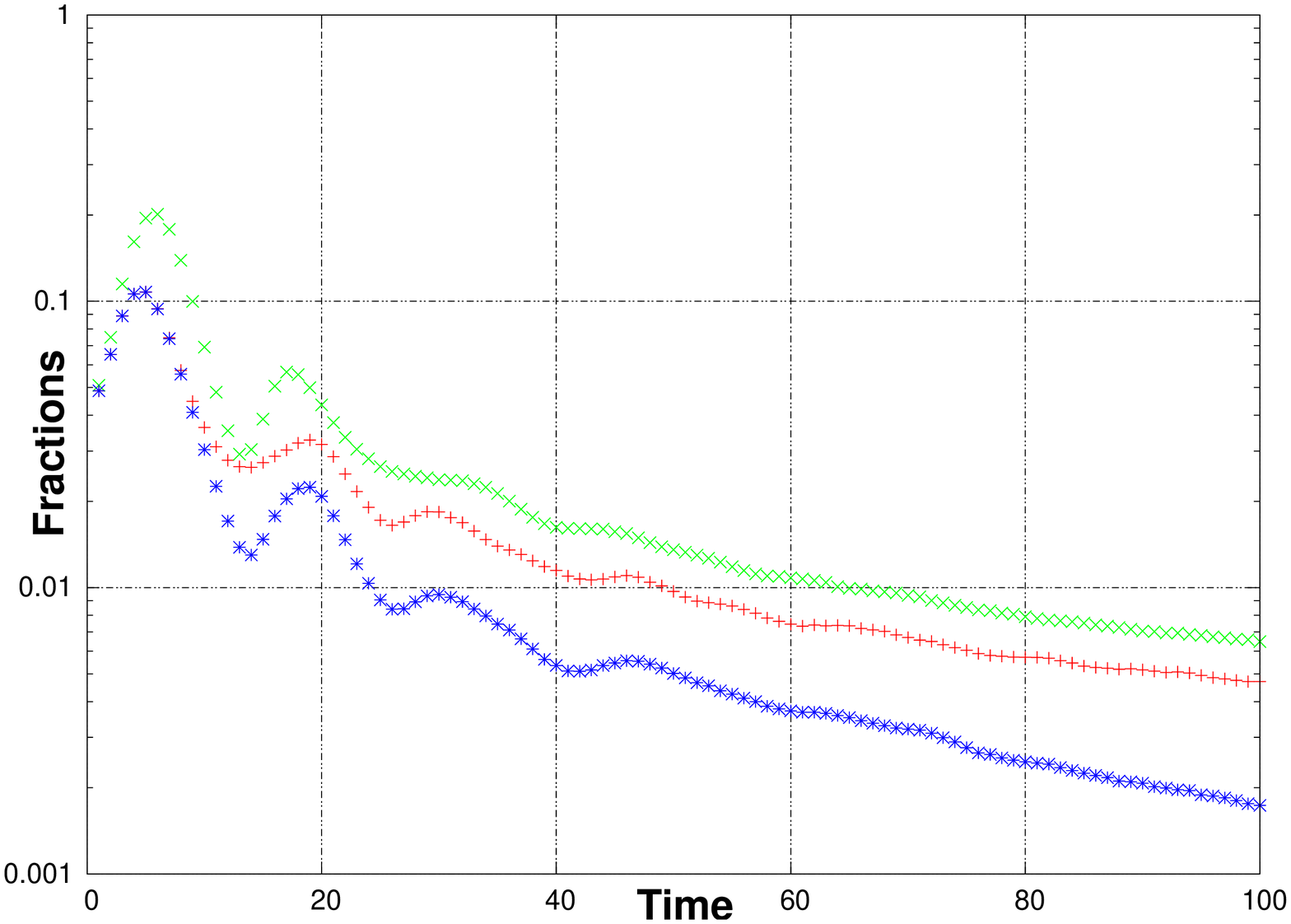}
\includegraphics[width=0.23\textwidth,angle=0]{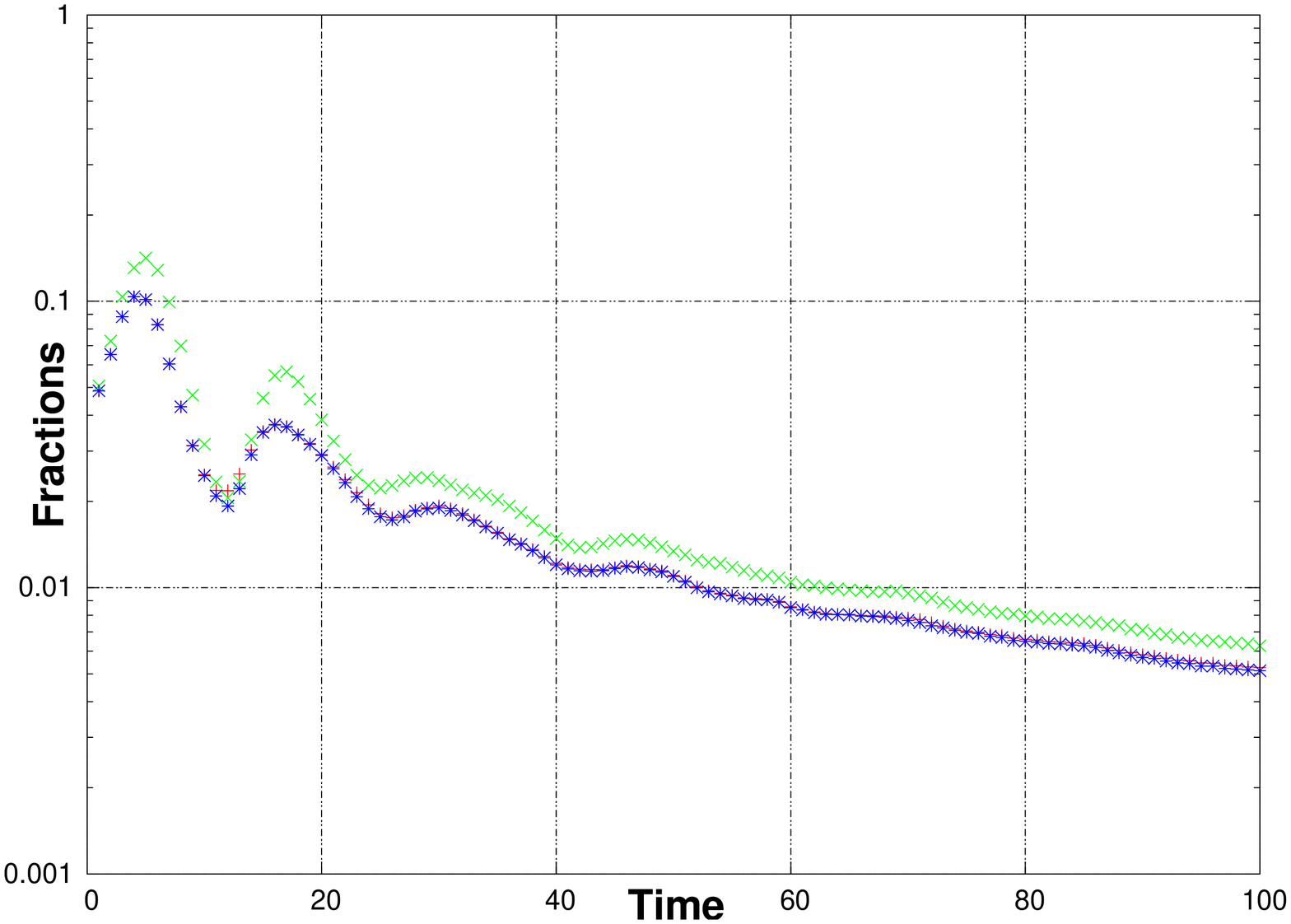}
\end{center} 
\caption{The total physical volume of the simulation box occupied by
  Higgs strings (green), axion strings (red), and their bound states
  (blue).  The left panels refer to local-global networks, while the
  right ones to local-local networks starting from the same initial
  conditions as the local-global ones. Figure taken from
  Ref.~\cite{Rajantie:2007hp}.
  \label{reverse_problem}} 
\end{figure}

Thus, different components of the $(p,q)$ state are expected to
exhibit different types of long-range interactions.  The evolution of
the string networks suggested that the long-range interactions have a
much more important r\^ole in the network evolution than the formation
of bound states. In the local-global networks the bound states tend to
split as a result of the long-range interactions, resulting in two
networks that evolve almost independently. The formation of
short-lived bound states and their subsequent splitting only increases
the small-scale {\sl wiggliness} of the local strings. In the case of
a local-local network, the absence of long-range interactions allows
the bound states to be much longer-lived and significantly influences
the evolution of the string network~\cite{Rajantie:2007hp}.  The most
convincing evidence comes from analysing the reverse
problem~\cite{Rajantie:2007hp}, namely that of a bound state splitting
as a result of the long-range interactions between strings, presented
in Fig.~\ref{reverse_problem}. Only in the absence of long-range
interactions the strings remain in the $\left(1,1\right)$ state
throughout their entire evolution.

\begin{figure}[htbp] 
  \begin{center} 
    \includegraphics[width=0.23\textwidth,angle=0]{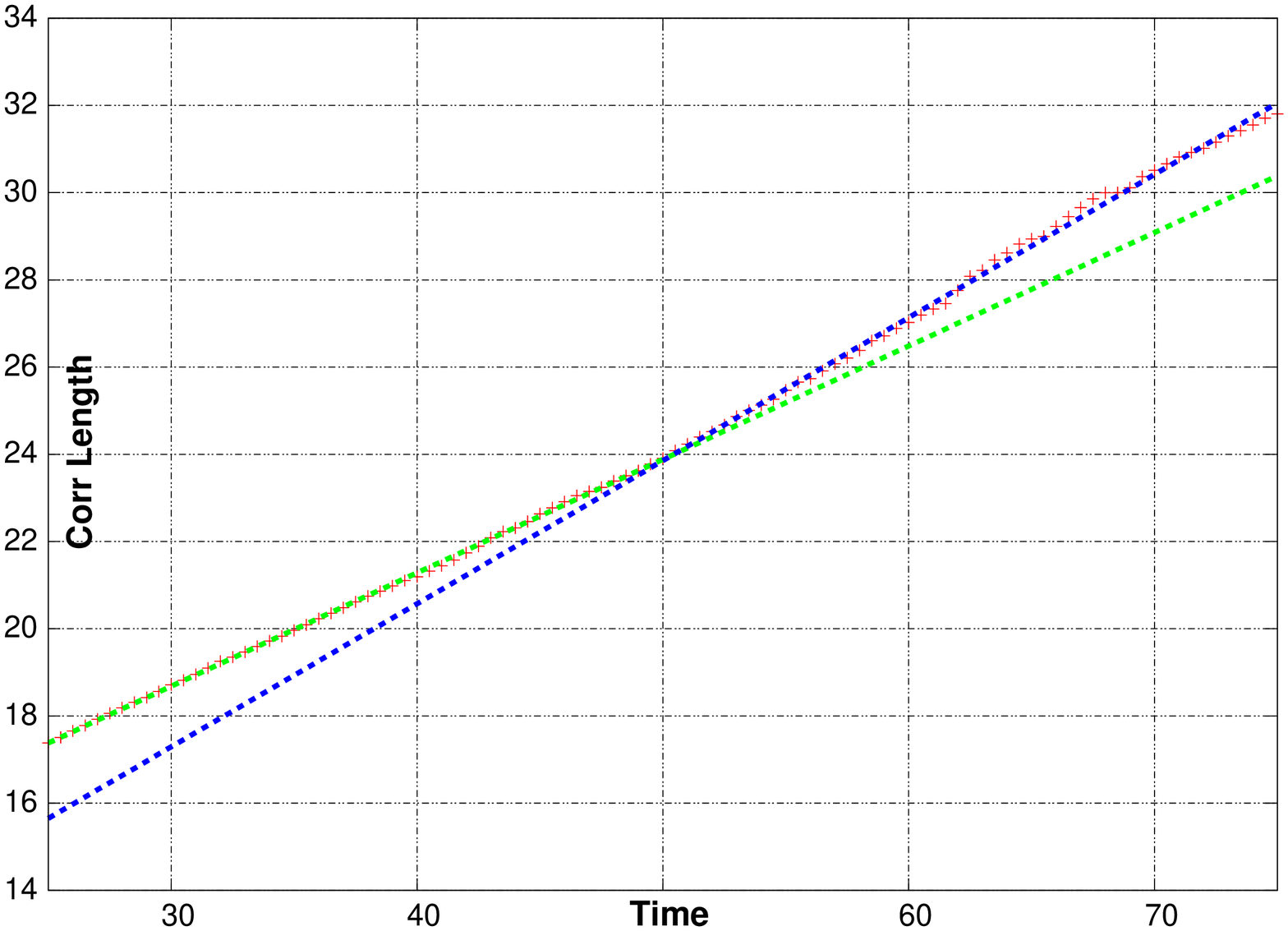}
    \includegraphics[width=0.23\textwidth,angle=0]{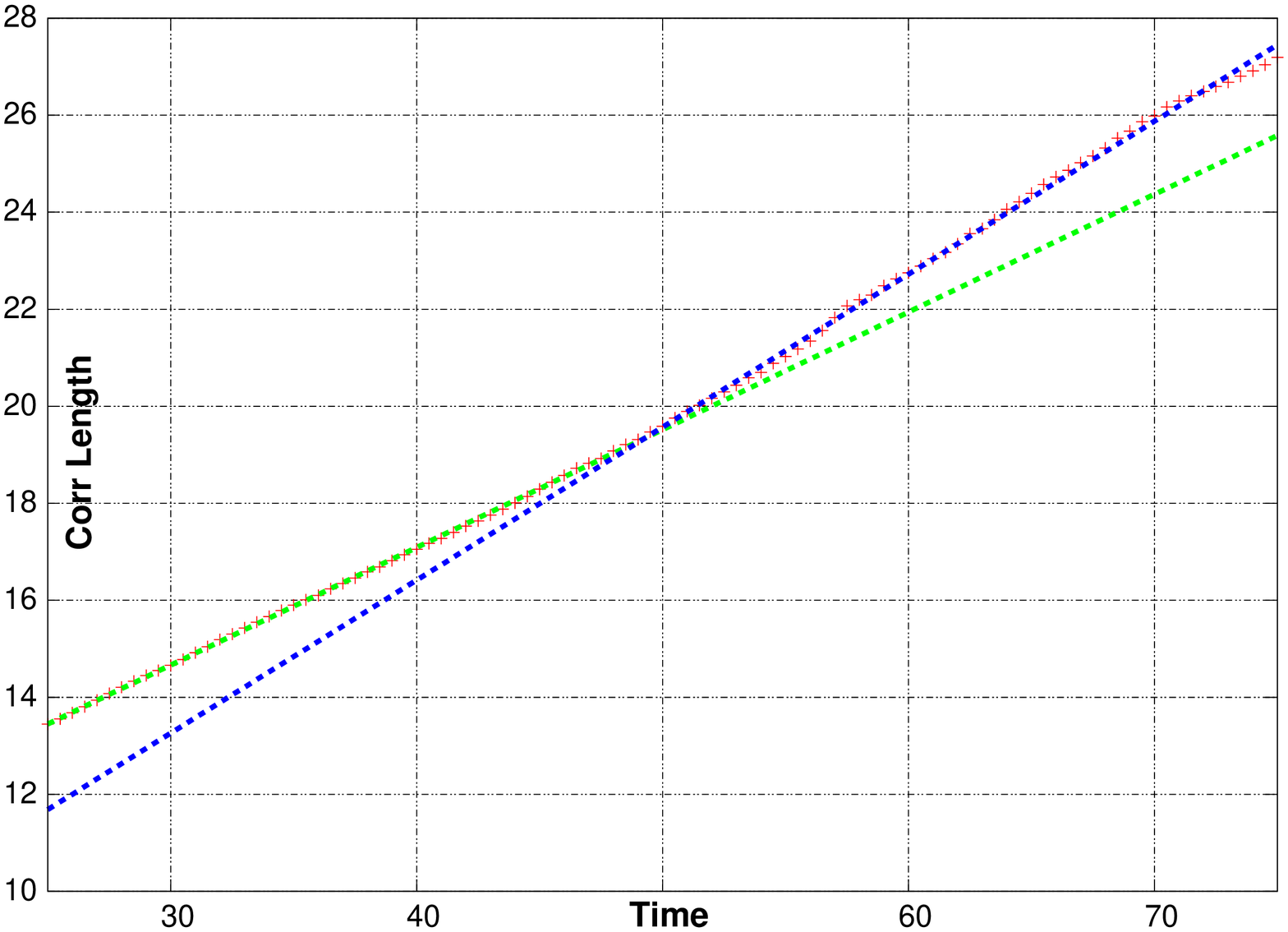}
    \includegraphics[width=0.23\textwidth,angle=0]{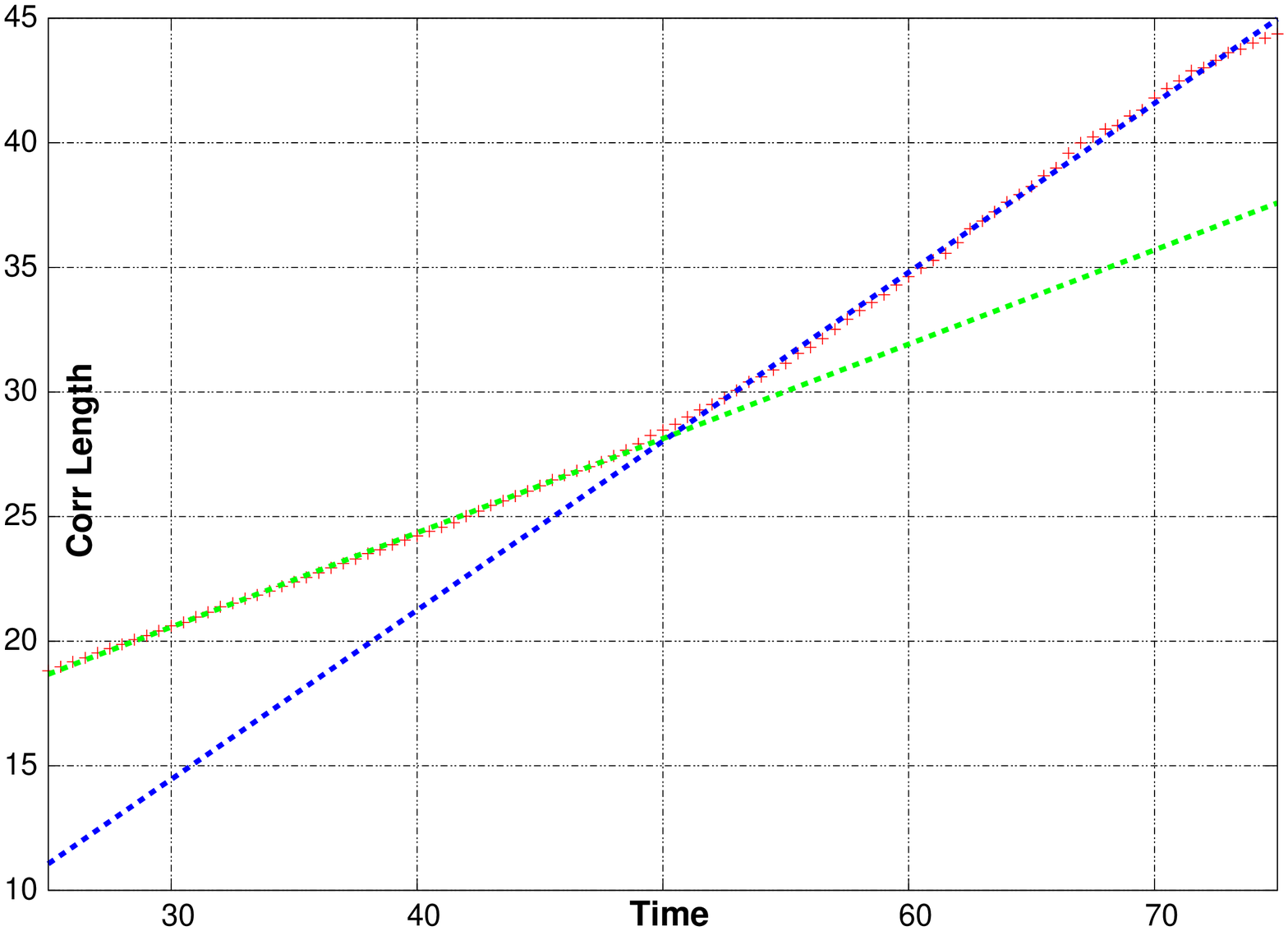}
    \caption{The Higgs (left), axion (middle), and bound state (right)
      string correlation length as a function of time. The network is
      a local-global one. The data and linear fits for the two regimes
      are shown. Figure taken from Ref.~\cite{Sakellariadou:2008ay}.
      \label{LG_High}} 
  \end{center} 
\end{figure}

\begin{figure}[htbp]
  \begin{center} 
    \includegraphics[width=0.23\textwidth,angle=0]{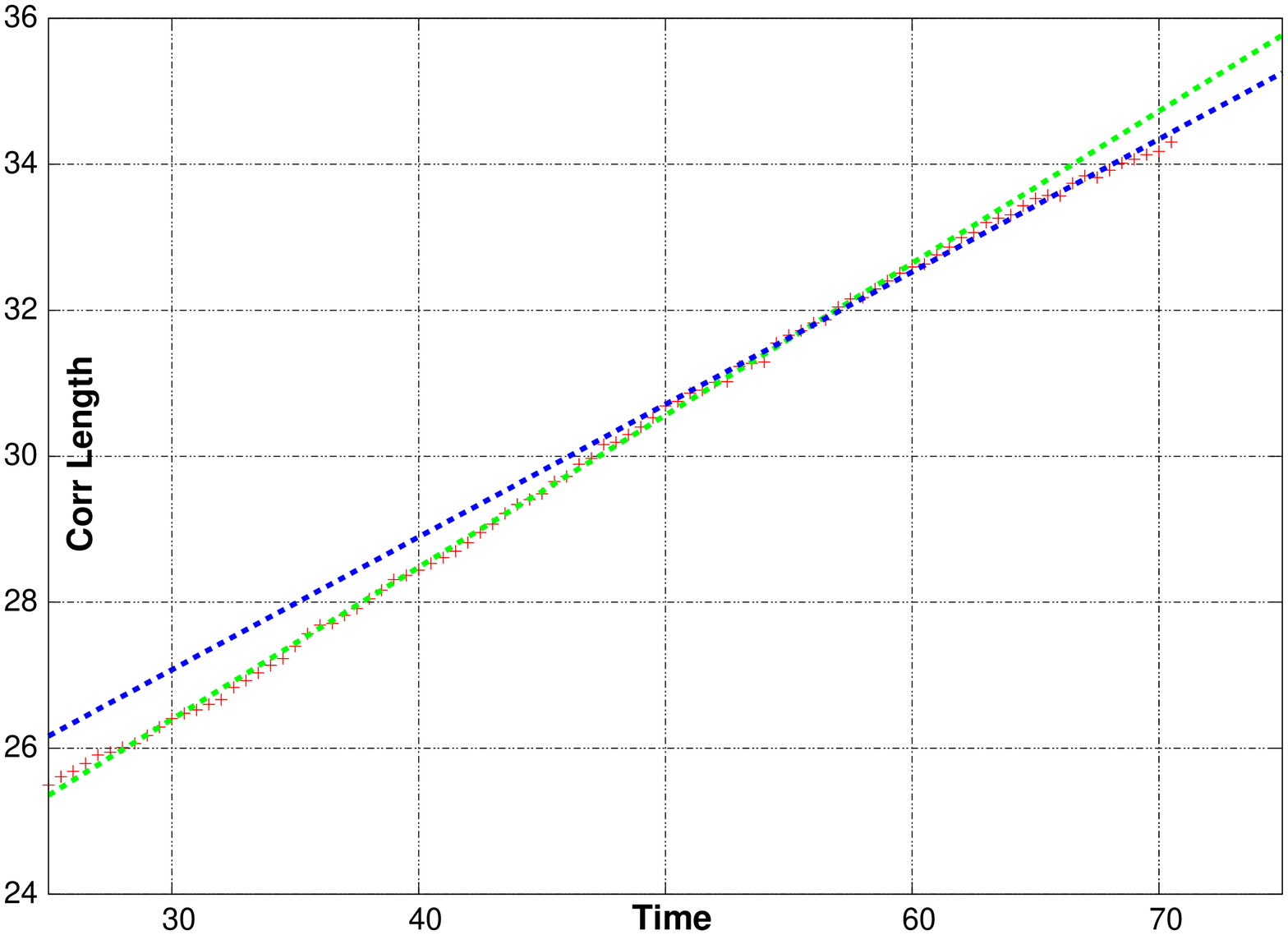}
    \includegraphics[width=0.23\textwidth,angle=0]{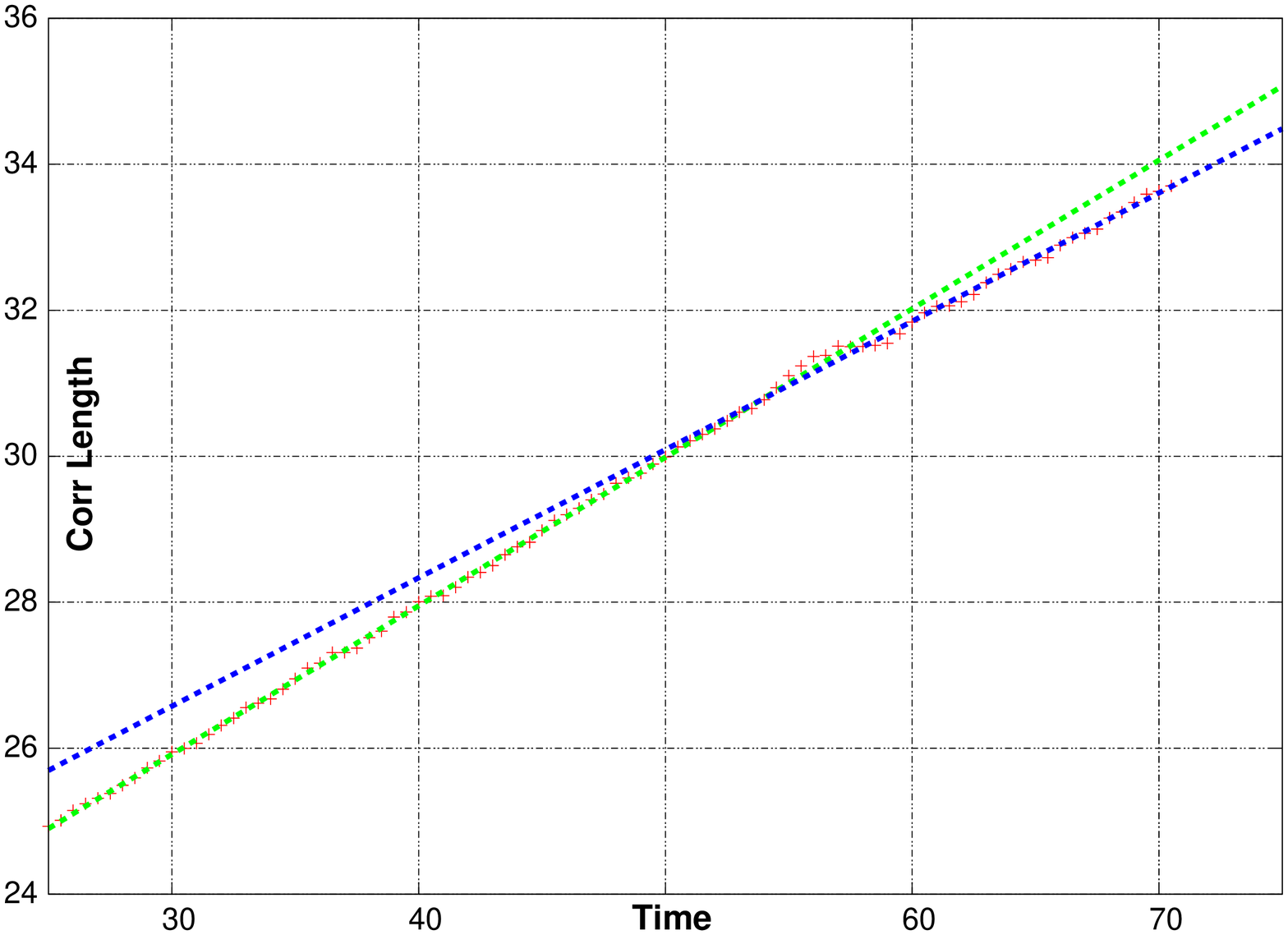}
    \includegraphics[width=0.23\textwidth,angle=0]{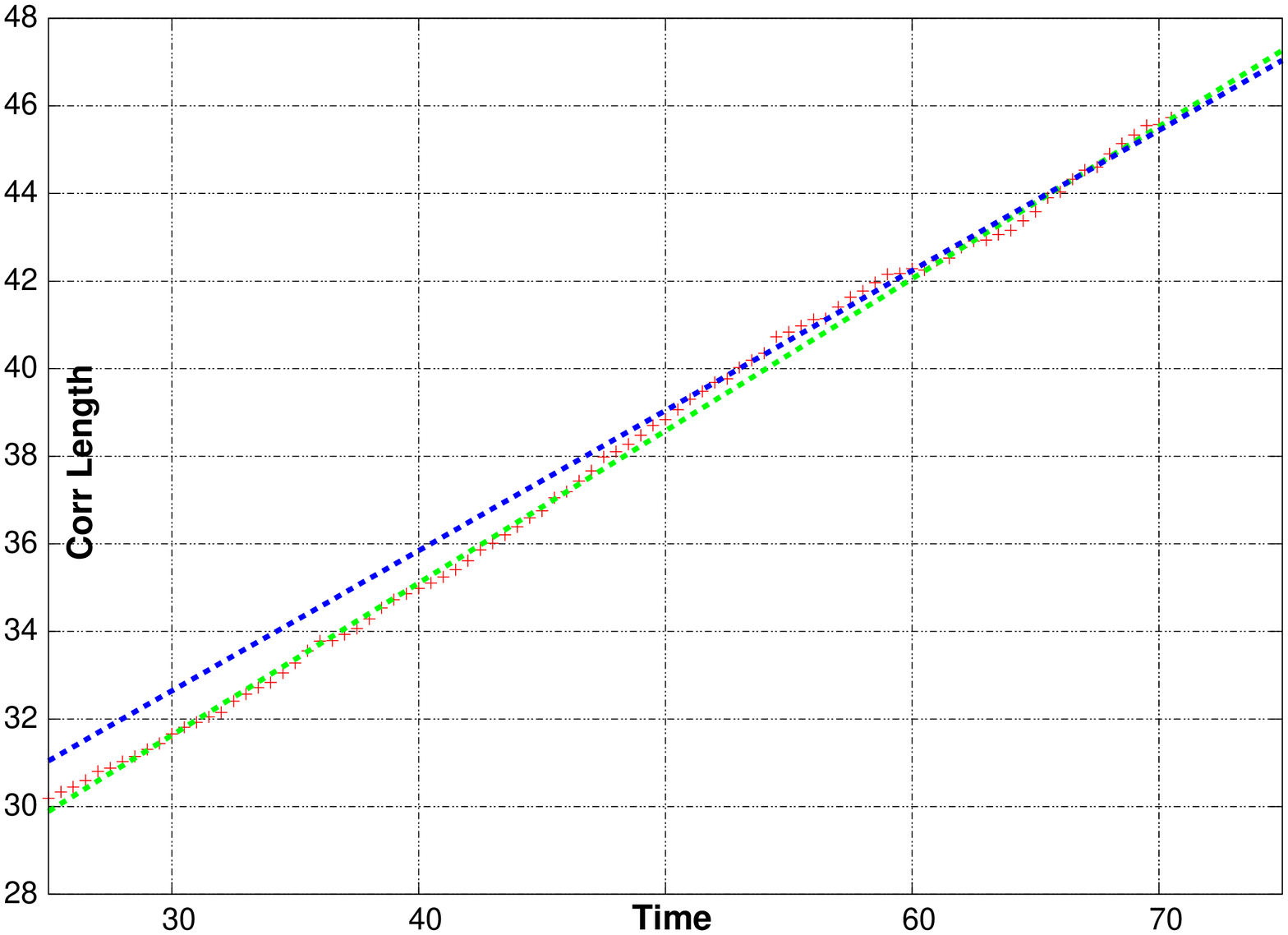}
    \caption{The same as Fig.~\ref{LG_High}, but for a local-local
      network. Figure taken from Ref.~\cite{Sakellariadou:2008ay}.
      \label{LL_High}} 
  \end{center} 
\end{figure}

Let us now investigate more thoroughly the issue of scaling.  The
evolution of F-, D-strings and their bound states is a rather
complicated problem, which necessitates both numerical as well as
analytical investigations. As I have already mentioned, junctions may
prevent the network from achieving a scaling solution, invalidating
the cosmological model leading to their formation.  Following the
approach of Ref.~\cite{Rajantie:2007hp}, numerical
simulations~\cite{Sakellariadou:2008ay}, achieving control over the
initial population of bound states, found clear evidence for scaling
of all three components --- $p$ F-strings, $q$ D-strings and their
$(p,q)$ bound states --- of the network, independently of the chosen
initial configurations, while they concluded that the existence of
bound states effects the evolution of the network.
In Fig.~\ref{LG_High} we show the string correlation length for the
Higgs and axion fields, as well as for their bound states, as a
function of time. The initial configuration is a local-global network
with a large amount of bound states.  
The corresponding plots for local-local networks are drawn in 
Fig.~\ref{LL_High}.
Clearly, there is convincing evidence for scaling of the three
components of the network for both networks. This scaling is
characterised with a distinct change of the correlation length slope
during the network evolution. Note that the result holds even in the
case of networks with small amounts of bound states.

Moreover, these numerical experiments have shown that for $(p,q)$
strings there is a supplementary energy loss mechanism, in addition to
the chopping off of loops; it is this new mechanism that allows the
network to scale. More precisely, the additional energy loss mechanism
is the formation of bound states, whose length increases, lowering the
overall energy of the network.

\section{Cosmic Superstrings: A window into String Theory} 

Cosmic superstrings have gained a lot of interest, the main reason
being that they can offer a large (and possibly unique) window into
string theory, and in particular shed some light on the appropriate
(if any) stringy description of the Universe. Since they interact with
Standard Model particles only via gravity,  their detection involves
their gravitational interactions.  Cosmic superstrings, in an analogy
to their solitonic analogues, can lead to a variety of astrophysical
signatures, like gravitational waves, ultra high energy cosmic rays,
and gamma ray bursts.

At this point, let me however emphasise that given the complexity of
the dynamics of a cosmic string network, which we certainly do not
fully understand, and the model-dependent initial configuration, any
theoretical estimations of the observational signatures of cosmic
superstrings have to be taken with caution. Note that even the
superstring tension depends on the considered inflationary scenario
within a particular brane-world cosmological model.

Gravitational waves is one of the main explored
avenues~\cite{Damour:2004kw}, in which case three channels of emission
have been identified. Radiation can be emitted by cusps, kinks, and/or
from the reconnection process itself.  Cusps, where momentarily the
string moves relativistically, have played a crucial r\^ole in
discussing the radiation emitted from (ordinary) cosmic strings
\footnote{Even though one has to keep in mind that the number of cusps
in a realistic cosmi string network has not been estimated, while
preliminary numerical studies indicate that it may be rather
low~\cite{christophe-mairi}.}.  Kinks, resulting from cosmic string
collisions and subsequent reconnection, are basically replaced in the
case of cosmic superstrings by junctions.  Finally, the radiation
emitted from the reconnection process itself, which a sub-dominant
process in the case of cosmic strings, may not be negligible in the
case of cosmic superstrings because of the small reconnection
probability ${\cal P}$.  In the scaling regime, the density of long
strings goes like $1 /{\cal P}$~\cite{Sakellariadou:2004wq}, implying
that the number of reconnection attempts goes like $1/{\cal P}^2$, and
hence the number of successful reconnections is approximately $1/{\cal
P}$. Very recently, the gravitational waveform produced by cosmic
superstring reconnections has been calculated~\cite{mx}. Comparing the
obtained result to the detection threshold for current and future
gravitational wave detectors, it was concluded~\cite{mx} that neither
bursts nor the stochastic gravitational background, produced during
the cosmic superstring reconnection process, would be detectable by
Advanced LIGO.  Thus, the most relevant process for gravitational
waves emitted from cosmic superstrings turns out to be through their
cusps. Hence, one should estimate the abundancy of cusps in cosmic
superstrings with junctions.

Following simple geometric arguments, it has been recently
shown~\cite{awsm} that strings ending on D-branes can indeed lead to
cusps, in an analogous way as cusps in ordinary cosmic strings. In
particular, cusps would be a generic feature of an F-string ending on
two (parallel and stationary) D-strings.  Hence, pairs of FD-string
junctions, such as those that they would form after intercommutations of F- and
D-strings, generically contain cusps.  This result opens up a new
energy loss mechanism for the network, in addition to the formation
and subsequent decay of closed loops and the formation of bound
states~\cite{Sakellariadou:2008ay}.  Phenomenological consequences of
cusps from junctions on cosmic superstrings will be most
significant at early times, namely towards the end of brane inflation, since
then the typical separation of heavy strings is small as compared to
the length of F-strings stretched between them~\cite{awsm}.  

\section{Cosmic Superstring Thermodynamics}
One has to extend previous studies of string thermodynamics
in the case of cosmic superstring networks, characterised by the
existence of $(p,q)$ bound states and different string tensions.
Recently, the Hagedorn transition of strings with junctions has been
investigated~\cite{smsj}, in the context of a simple model with three
different types and tensions of string, following an effective field
theory approach. More precisely, the authors of Ref.~\cite{smsj}
translated the thermodynamics of string networks with junctions into
the thermodynamics of a set of interacting dual fields. Thus, the
Hagedorn transition of the strings becomes a transition of the fields.

In this approach, the equilibrium statistical mechanics of cosmic
superstring networks have been studied~\cite{smsj}, by extending known
methods for describing quark deconfinement. It was found~\cite{smsj}
that as the system is heated, the lightest strings are the first ones
to undergo a Hagedorn transition; the existence of junctions does not
affect the occurrence of the transition.  The system is also
characterised by a second, higher, critical temperature above which
long string modes of all tensions and junctions, do exist.  The
existence of multiple tensions indicates the appearance of multiple
Hagedorn transitions.

\section{Conclusions}
In these lectures, I have summarised our current understanding on the
physics of cosmic strings and cosmic superstrings. I have discussed
their formation, evolution, statistical mechanics and
astrophysical/cosmological consequences. This is a topic of active
research at present, relating fundamental theoretical ideas with
experimental and observational facts.

On the one hand, any successful cosmological scenario, such as the
inflationary paradigm, must be inspired from a fundamental theory. On
the other hand, any successful high energy physics theory, such as
string theory or supersymmetric grand unified theories, must be tested
against data; the only available laboratory for the required energy
scales, is indeed the early Universe.

Inflation within brane-world cosmological models leads naturally to
cosmic superstrings.  Inflation within supersymmetric grand unified
theories leads generically to cosmic strings, the solitonic analogues
of cosmic superstrings. The study of these objects is interesting by
itself. In addition, cosmic (super)strings may provide an explanation
for the origin of a variety of astrophysical/cosmological
observations; they may also offer a test (often a unique one) of
fundamental theories of physics, thus shedding some light about the
appropriate stringy description of the Universe.

 \vskip1.truecm
\noindent
{\bf Acknowledgments}\\ \\ It is a pleasure to thank the organisers of
the ESF Summer School in High Energy Physics and Astrophysics ``Theory
and Particle Physics: the LHC perspective and beyond'', which took
place  at the Carg\`ese Institute of Scientific Studies in Corsica,
for inviting me to present these lectures in such a beautiful and
stimulating environment.

\end{document}